\documentclass[longauth]{aa}  

\usepackage{subcaption}
\usepackage{graphicx}
\usepackage{comment}
\usepackage[dvipsnames]{xcolor}
\usepackage{multirow}
\usepackage{hhline}
\usepackage{ulem}

\usepackage{txfonts}
\usepackage{hyperref}
\hypersetup{colorlinks,linkcolor={blue},citecolor={blue}}

\def\clj{{CL~J1226.9+3332}}

\begin{document}

   \title{
   Thermal Sunyaev-Zel'dovich effect at the core of \clj\ revealed by NOEMA
   }
   \titlerunning{Thermal Sunyaev-Zel'dovich effect at the core of \clj\ revealed by NOEMA}

  \author{M.~Muñoz-Echeverr\'ia\inst{\ref{IRAP}}\thanks{miren.munoz-echeverria@irap.omp.eu} 
    \and J.-F.~Mac\'ias-P\'erez\inst{\ref{LPSC}} 
    \and R.~Neri\inst{\ref{IRAMF}}
    \and E.~Pointecouteau\inst{\ref{IRAP}}
    \and R.~Adam \inst{\ref{OCA}}
    \and  P.~Ade \inst{\ref{Cardiff}}
    \and  H.~Ajeddig \inst{\ref{CEA}}
    \and  S.~Amarantidis \inst{\ref{IRAME}}
    \and  P.~Andr\'e \inst{\ref{CEA}}
    \and  H.~Aussel \inst{\ref{CEA}}
    \and  A.~Beelen \inst{\ref{LAM}}
    \and  A.~Beno\^it \inst{\ref{Neel}}
    \and  S.~Berta \inst{\ref{IRAMF}}
    \and  M.~B\'ethermin \inst{\ref{Strasbourg}}
    \and  A.~Bongiovanni \inst{\ref{IRAME}}
    \and  J.~Bounmy \inst{\ref{LPSC}}
    \and  O.~Bourrion \inst{\ref{LPSC}}
    \and  M.~Calvo \inst{\ref{Neel}}
    \and  A.~Catalano \inst{\ref{LPSC}}
    \and  D.~Ch\'erouvrier \inst{\ref{LPSC}}
    \and  U.~Chowdhury \inst{\ref{Neel}}        
    \and  M.~De~Petris \inst{\ref{Roma}}
    \and  F.-X.~D\'esert \inst{\ref{IPAG}}
    \and  S.~Doyle \inst{\ref{Cardiff}}
    \and  E.~F.~C.~Driessen \inst{\ref{IRAMF}}
    \and  G.~Ejlali \inst{\ref{Teheran}}
    \and  A.~Ferragamo \inst{\ref{Roma}}
    \and  A.~Gomez \inst{\ref{CAB}} 
    \and  J.~Goupy \inst{\ref{Neel}}
    \and  C.~Hanser \inst{\ref{CPPM}}
    \and  S.~Katsioli \inst{\ref{AthenObs}, \ref{AthenUniv}}
    \and  F.~K\'eruzor\'e \inst{\ref{Argonne}}
    \and  C.~Kramer \inst{\ref{IRAMF}}
    \and  B.~Ladjelate \inst{\ref{IRAME}} 
    \and  G.~Lagache \inst{\ref{LAM}}
    \and  S.~Leclercq \inst{\ref{IRAMF}}
    \and  J.-F.~Lestrade \inst{\ref{LERMA}}
    \and  S.~C.~Madden \inst{\ref{CEA}}
    \and  A.~Maury \inst{\ref{Barcelona1}, \ref{Barcelona2}, \ref{CEA}}
    \and  F.~Mayet \inst{\ref{LPSC}}
    \and  A.~Monfardini \inst{\ref{Neel}}
    \and  A.~Moyer-Anin \inst{\ref{LPSC}}
    \and  I.~Myserlis \inst{\ref{IRAME}}
    \and  A.~Paliwal \inst{\ref{Roma2}}
    \and  L.~Perotto \inst{\ref{LPSC}}
    \and  G.~Pisano \inst{\ref{Roma}}
    \and  N.~Ponthieu \inst{\ref{IPAG}}
    \and  V.~Rev\'eret \inst{\ref{CEA}}
    \and  A.~J.~Rigby \inst{\ref{Leeds}}
    \and  A.~Ritacco \inst{\ref{LPSC}}
    \and  H.~Roussel \inst{\ref{IAP}}
    \and  F.~Ruppin \inst{\ref{IP2I}}
    \and  M.~S\'anchez-Portal \inst{\ref{IRAME}}
    \and  S.~Savorgnano \inst{\ref{LPSC}}
    \and  K.~Schuster \inst{\ref{IRAMF}}
    \and  A.~Sievers \inst{\ref{IRAME}}
    \and  C.~Tucker \inst{\ref{Cardiff}}
    \and  R.~Zylka \inst{\ref{IRAMF}}  
    \and I.~Bartalucci \inst{\ref{INAFmilano}}
    \and  J.-B.~Melin \inst{\ref{IRFU}}
    \and  G.~W.~Pratt \inst{\ref{CEA}}
  }
  
  \authorrunning{M.~Muñoz-Echeverr\'ia et al.}

 \institute{
    IRAP, CNRS, Université de Toulouse, CNES, UT3-UPS, (Toulouse), France
    \label{IRAP}
    \and
    Université Grenoble Alpes, CNRS, LPSC-IN2P3, 53, avenue
    des Martyrs, 38000 Grenoble, France
    \label{LPSC}      
    \and
    Institut de Radioastronomie Millim\'etrique (IRAM), 300 rue de la Piscine, 38406 Saint-Martin-d'Hères, France
    \label{IRAMF}
    \and
    Universit\'e C\^ote d'Azur, Observatoire de la C\^ote d'Azur, CNRS, Laboratoire Lagrange, France 
    \label{OCA}
    \and
    School of Physics and Astronomy, Cardiff University, Queen’s Buildings, The Parade, Cardiff, CF24 3AA, UK 
    \label{Cardiff}
    \and
    Université Paris Cité, Université Paris-Saclay, CEA, CNRS, AIM, F-91191 Gif-sur-Yvette, France
    \label{CEA}
    \and
    Institut de Radioastronomie Millim\'etrique (IRAM), Avenida Divina Pastora 7, Local 20, E-18012, Granada, Spain
    \label{IRAME}     
    \and        
    Aix Marseille Univ, CNRS, CNES, LAM (Laboratoire d'Astrophysique de Marseille), Marseille, France
    \label{LAM}
    \and
    Institut N\'eel, CNRS, Universit\'e Grenoble Alpes, France
    \label{Neel}
    \and
    Universit\'e de Strasbourg, CNRS, Observatoire astronomique de Strasbourg, UMR 7550, 67000 Strasbourg, France
    \label{Strasbourg}
    \and 
    Dipartimento di Fisica, Sapienza Universit\`a di Roma, Piazzale Aldo Moro 5, I-00185 Roma, Italy
    \label{Roma}
    \and
    Univ. Grenoble Alpes, CNRS, IPAG, 38000 Grenoble, France 
    \label{IPAG}
    \and
    Institute for Research in Fundamental Sciences (IPM), School of Astronomy, Tehran, Iran
    \label{Teheran}
    \and
    Centro de Astrobiolog\'ia (CSIC-INTA), Torrej\'on de Ardoz, 28850 Madrid, Spain
    \label{CAB}
    \and
    Aix Marseille Univ, CNRS/IN2P3, CPPM, Marseille, France
    \label{CPPM}
    \and
    National Observatory of Athens, Institute for Astronomy, Astrophysics, Space Applications and Remote Sensing, Ioannou Metaxa
    and Vasileos Pavlou GR-15236, Athens, Greece
    \label{AthenObs}
    \and
    Department of Astrophysics, Astronomy \& Mechanics, Faculty of Physics, University of Athens, Panepistimiopolis, GR-15784
    Zografos, Athens, Greece
    \label{AthenUniv}
    \and
    High Energy Physics Division, Argonne National Laboratory, 9700 South Cass Avenue, Lemont, IL 60439, USA
    \label{Argonne}
    \and  
    LUX, Observatoire de Paris, PSL Research University, CNRS, Sorbonne Universit\'e, UPMC, 75014 Paris, France
    \label{LERMA}
    \and
    Institute of Space Sciences (ICE), CSIC, Campus UAB, Carrer de Can Magrans s/n, E-08193, Barcelona, Spain
    \label{Barcelona1}
    \and
    ICREA, Pg. Lluís Companys 23, Barcelona, Spain
    \label{Barcelona2}
    \and
    Dipartimento di Fisica, Universit\`a di Roma ‘Tor Vergata’, Via della Ricerca Scientifica 1, I-00133 Roma, Italy      
    \label{Roma2}
    \and
    School of Physics and Astronomy, University of Leeds, Leeds LS2 9JT, UK
    \label{Leeds}
    \and
    Laboratoire de Physique de l’\'Ecole Normale Sup\'erieure, ENS, PSL Research University, CNRS, Sorbonne Universit\'e, Universit\'e de Paris, 75005 Paris, France 
    \label{ENS}
    \and
    INAF-Osservatorio Astronomico di Cagliari, Via della Scienza 5, 09047 Selargius, IT
    \label{INAF}
    \and    
    Institut d'Astrophysique de Paris, CNRS (UMR7095), 98 bis boulevard Arago, 75014 Paris, France
    \label{IAP}
    \and
    University of Lyon, UCB Lyon 1, CNRS/IN2P3, IP2I, 69622 Villeurbanne, France
    \label{IP2I}
    \and
    University Federico II, Naples, Italy
    \label{Naples}    
    \and
     INAF, IASF-Milano, via A. Corti 12, 20133, Milano, Italy
    \label{INAFmilano}
    \and
     IRFU, CEA, Universit\'e Paris-Saclay, 91191, Gif-sur-Yvette, France
    \label{IRFU}
   }

   \date{Received --------; accepted -------}

 
  \abstract
  {We present first detailed maps of the thermal Sunyaev-Zel'dovich (tSZ) effect on a $z=0.89$ cluster with the NOrthern Extended Millimeter Array (NOEMA). The high sensitivity of these observations enabled the effective identification and removal of the millimetre-wave sources contaminating the tSZ signal, thus isolating the influence of the hot electron gas of the cluster on the cosmic microwave background radiation from other emissions. The tSZ observed with success by NOEMA was modelled together with previous single-dish observations (IRAM 30-metre, Green Bank Telescope, and Caltech Sub-millimeter Observatory) to obtain the first core-to-outskirts (from $\sim$~15 to $\sim$~1500~kpc) pressure profile reconstruction on such a high-redshift galaxy cluster. NOEMA observations with a high angular resolution have shown that the pressure profile is flat in the core of the cluster. These observations confirm the disturbed nature of \clj\ and map for the first time the distribution of its thermal gas at arcsecond scales in the environments of the central cluster galaxy. Our results showcase the excellent capabilities of NOEMA to complement and enhance the data provided by other millimetre-wave instruments in resolving the core of high-redshift clusters via tSZ emission.
  }


   \maketitle

\section{Introduction}
\label{sec:intro}

The understanding of galaxy cluster properties and their evolution through cosmic time contributes to our comprehension of the structure formation in the Universe \citep{2011ARA&A..49..409A}. Mainly driven by gravitational forces, the collapse and accretion of matter are also sensitive to the baryonic physics of the hot ionised gas in the intracluster medium \citep[ICM,][]{2014MNRAS.439.2485C, 2023A&A...678A.109A}. 

The ICM of clusters can be observed in X-rays or through the imprint of the thermal Sunyaev-Zel'dovich (tSZ) effect \citep{1972CoASP...4..173S} on the cosmic microwave background radiation (CMB). Current galaxy cluster observations at millimetre wavelengths with single-dish telescopes are limited at high angular resolution, which prevents us from resolving the structures in the core of clusters. The 
New IRAM KID Array 2 \citep[NIKA2,][]{adam_nika2_2018, NIKA2-electronics,calvo} at the IRAM 30-metre telescope and the Multiplexed SQUID/TES Array at Ninety Gigahertz 2
\citep[MUSTANG2,][]{dicker_mustang2_2014} on the 100-metre Green Bank Telescope currently offer the best capabilities. NIKA2 maps the sky at 150 GHz and 260 GHz with $17.6^{^{\prime\prime}}$ and $11.1^{^{\prime\prime}}$ full width at half maximum (FWHM) angular resolutions, respectively \citep{2020A&A...637A..71P}. MUSTANG2 has a $9^{^{\prime\prime}}$ FWHM beam at 90 GHz. Increasing the resolution is currently only possible with millimetre-wave
interferometers such as the Atacama Large Millimeter/submillimeter Array (ALMA) and the NOrthern Extended Millimeter Array \citep[NOEMA,][]{2023pcsf.conf..308N}.  \\

Historically, radio interferometers have had an incomparable importance in the detection of clusters of galaxies through the tSZ effect. To name a few, the Very Large Array \citep[VLA,][]{Moffet&Birkinshaw1989}, the Very Small Array \citep[VSA,][]{vsa}, the Ryle telescope \citep{jones1993,grainge1993}, the Berkeley-Illinois-Maryland Array \citep[BIMA,][]{carlstrom1996}, the Sunyaev-Zel'dovich Array \citep[SZA,][]{muchovej}, or the Arcminute MicroKelvin Interferometer \citep[AMI,][]{zwart2008} imaged the tSZ signal of clusters for the first time \citep[see][for a review on the status of cluster observations through the tSZ effect at the beginning of the 21st century]{carlstrom2002}. Recently, the capabilities of the ALMA instrument (combined in some cases with data from the Atacama Compact Array; ACA) have led to interferometric tSZ observations of clusters that are able to resolve structures and merger shocks in high-redshift objects at arcsecond scales \citep{kitayama2016, basu2016, Di_Mascolo_2023, 2024A&A...682A...186L, 2023arXiv231006120V}. In the Northern Hemisphere, this will only be possible if
the tSZ effect induced by clusters can be observed with NOEMA.

In this paper, we present the observations of the \clj\ galaxy cluster with NOEMA, which allowed us the first detection of the tSZ effect with this facility. The high angular resolution tSZ data were used to reconstruct the distribution of the thermal pressure in the core of the cluster, and subsequently, they were combined with other millimetre-data-based pressure reconstructions to obtain the pressure profile of \clj\ with an unprecedented level of radial coverage for a cluster at redshift $z = 0.89$.

This paper is organised as follows. In Sect.~\ref{sec:target} we motivate and present the observations of \clj\ with NOEMA. The analysis of NOEMA data is described in Sect.~\ref{sec:dataanalysis}. In the same section, we present the detection of the tSZ effect in the NOEMA visibilities and image space map. A pressure profile model is fitted to the tSZ signal measured by NOEMA in Sect.~\ref{sec:pressure} and is then combined with previous pressure reconstructions to obtain a precise description of the thermal pressure distribution in \clj\ . Finally, the summary and conclusions are given in Sect.~\ref{sec:conclusion}. Throughout this work, we assume a flat $\Lambda$CDM cosmology with $H_{0} = 70 \hspace{2pt} \mathrm{km/s/Mpc}$ and $\Omega_{m,0} = 0.3$. Thus, at the cluster redshift, 1 arcmin corresponds to a physical distance of 466~kpc.

\section{Target and observations}
\label{sec:target}

Based on its specific characteristics and existing data, the \clj\  galaxy cluster was selected as an interesting target for the initial attempt of NOEMA to detect the tSZ effect.
This cluster has been widely observed and studied at different wavelengths \citep{2009ApJ...691.1337J, 2015ApJ...801...44Z, 2021NatAs...5..268D}. It is a hot and massive object \citep[see][and references therein]{2023A&A...671A..28M}, which ensures a prominent tSZ signal. According to X-ray observations \citep[][]{2007ApJ...659.1125M}, within an overdensity radius of 500 times the critical density, it reaches temperatures of 10.7~keV. In the millimetre domain, 
BIMA \citep{2001ApJ...551L...1J}, NIKA \citep{2015A&A...576A..12A}, \textit{Planck} \citep{2016A&A...594A..27P}, MUSTANG \citep{2017ApJ...838...86R}, Bolocam \citep{2017ApJ...838...86R}, and NIKA2 \citep{2023A&A...671A..28M} observations have contributed to the understanding of the distribution of the gas within the ICM of \clj\ at large scales ($>9^{\prime\prime}$).

Located at a redshift of $z = 0.89$, \clj\ is the cluster with the highest redshift of the NIKA2 SZ Large Programme \citep{2020EPJWC.22800017M, 2022EPJWC.25700038P} sample, and it is therefore one of the most compact sources because of its angular size. We know from previous observations that the tSZ signal of the cluster is contaminated by point sources in the field and that the core of the cluster is very likely dynamically disturbed \citep{2007ApJ...659.1125M,2017ApJ...838...86R}.
At this redshift, observations at arcsecond resolution are needed in order to map the distribution of the gas in the core. In addition, the NOEMA capabilities, which combine a high angular resolution and sensitivity, are uniquely suited to identifying the contaminating galaxies in the field. These aspects together make a compelling scientific case to be investigated with NOEMA.

\begin{figure}[h!]
    \centering
    \includegraphics[scale=0.25,trim={1cm 0cm 0cm 0cm}]{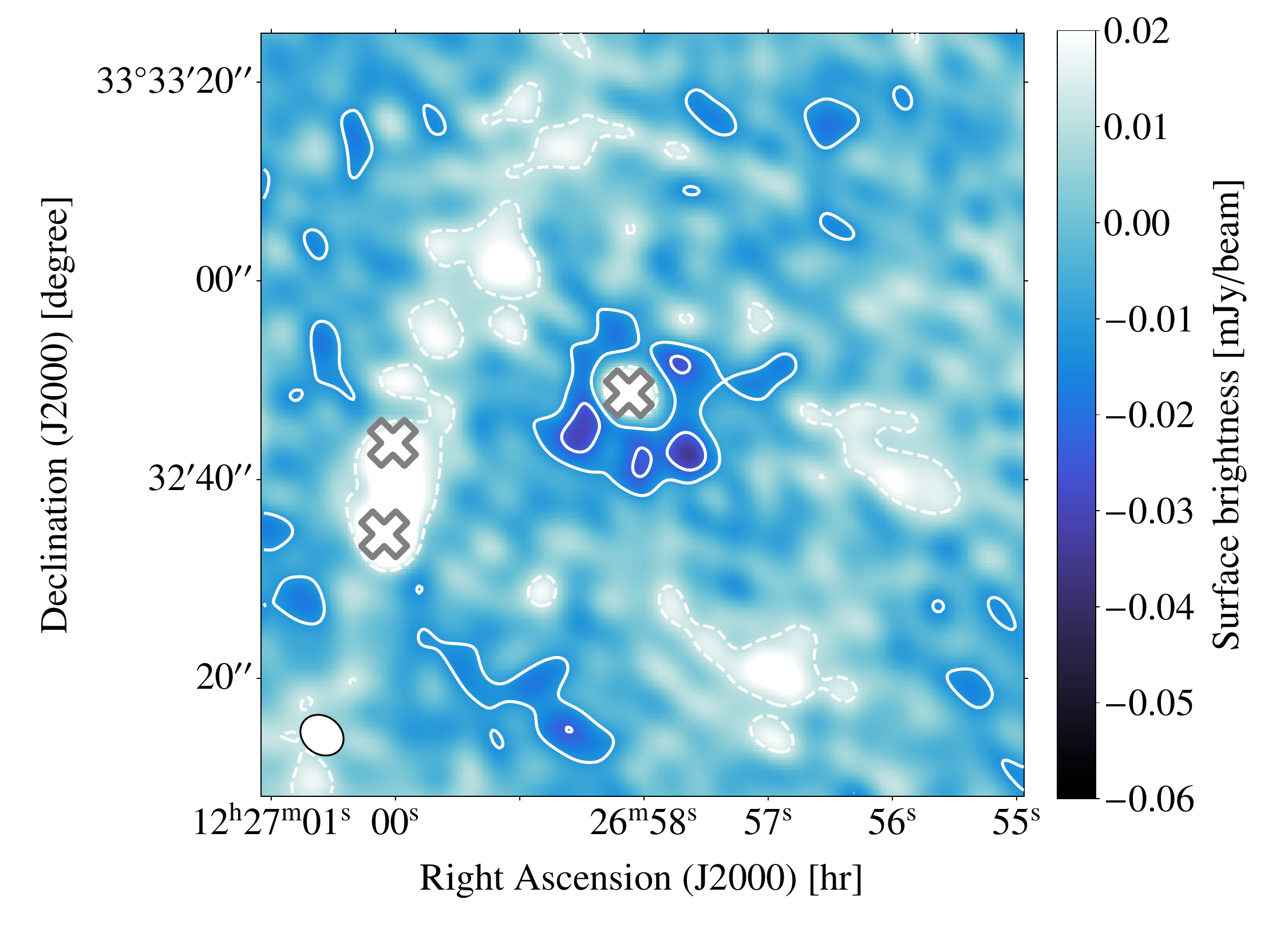}
    \includegraphics[scale=0.25, trim={1cm 0cm 0cm 0cm}]{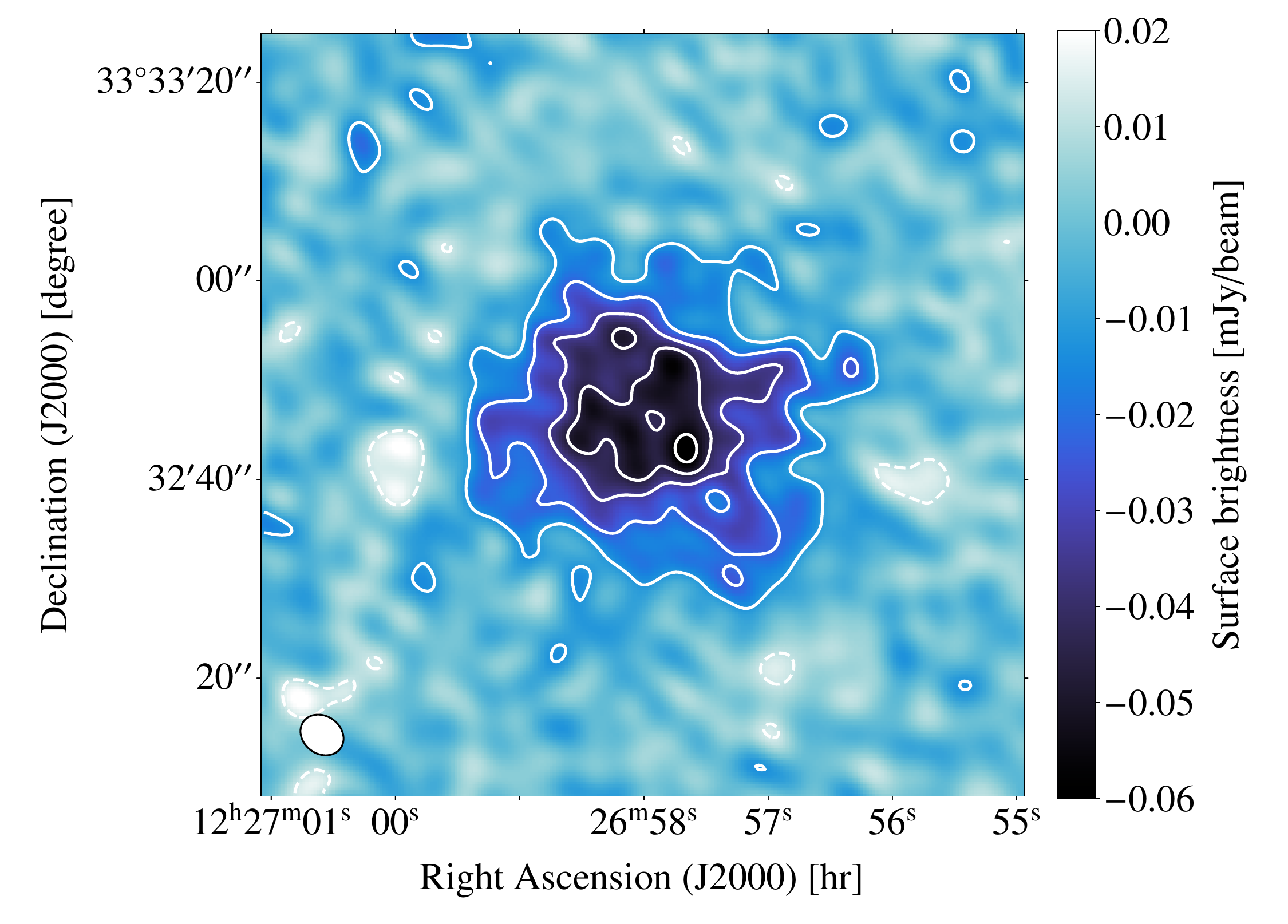}
    \caption{Top: Continuum NOEMA dirty map of \clj\ from the combination of point-source-contaminated LSB and USB data, merged without weighting for the tSZ effect spectral shape. The grey crosses indicate the positions of the three point sources identified from NOEMA data (Table~\ref{tab:firstfit}).
    Bottom: Point-source-subtracted continuum NOEMA map of \clj\ 
    cleaned with \texttt{CLARK} \citep{1980A&A....89..377C} and natural weighting. The data from different sidebands were combined as described in Sect.~\ref{sec:continuum}. The map is not corrected for primary beam attenuation. The white contours indicate $\pm 2\sigma = \pm 11.2 \; \mu$Jy~beam$^{-1}$ levels spaced by $2\sigma$. The size and orientation of the synthesized beam of NOEMA (4.6$^{\prime\prime}\,\times\,$3.8$^{\prime\prime}$ at PA $=54^\circ$) are indicated in the lower left corners. 
    }
    \label{fig:noemamaps}
\end{figure}

The tSZ signal is extended and weak. We therefore selected the most compact (D) configuration of NOEMA to maximise the sensitivity. The observations of the galaxy cluster were conducted from September 12, 2022, to May 29, 2023 in the single-field mapping mode using ten antennas. These observations amounted to 19.5 hours of on-source time. The phase-tracking centre was positioned on the core of the cluster (as defined by the X-ray peak) at (RA, Dec)$_{\mathrm{J2000}}$ = (12$^{\rm h}$26$^{\rm m}$58.00$^{\rm s}$, +33${^\circ}$32$'$46.68$^{\prime\prime}$). The correlator was set to observe with the lowest spectral resolution configuration ($2\times 10^{-3}$~GHz channels) and to cover the $70-78$~GHz and $86-94$~GHz frequency ranges in the lower and upper receiver sidebands (LSB and USB), respectively, and in both polarisations. Phase and amplitude calibrations of the observations were performed using J1310+323.
Flux calibration was performed using the IRAM models of the primary calibrators MWC~349 and LkH$\alpha$~101, whose 90~GHz flux densities (1.18 and 0.23~Jy, respectively) are continuously monitored, vary by less than 5\% over the period of our observations, and show stable millimetre-wave spectral indices (0.60 and 0.96, respectively). The top panel in Fig.~\ref{fig:noemamaps} shows the continuum dirty map of NOEMA \clj\ observations obtained from the concatenation of LSB and USB data.

\renewcommand{\arraystretch}{1.4} 
\begin{table*}[h!]
\caption{Best-fitting positions of the point sources and the negative CMB distortion fitted to the continuum NOEMA $uv$ table.}
    \centering
    \begin{tabular}{c c c c c c c }
    \hline \hline
         RA & Dec & Source  & \texttt{uv\_fit} function & LSB flux &  USB flux & FWHM  \\ 
         J2000 & J2000 & ~ &  ~ & [$\mu$Jy] & [$\mu$Jy] & [arcsec]\\ \hline
        12:26:58.12 (0.21) &  +33:32:48.7 (0.3) & PS9 & \texttt{point} & $72.7 \pm 9.5$ & $64.5 \pm 6.7$ & ~ \\
        12:27:00.09 (0.30) & +33:32:34.5 (1.0)& $\text{PS260-S}$ & \texttt{point} & $80.8  \pm 16.4$ & $121.8 \pm 14.9$ & ~ \\
        12:27:00.02 (0.60) & +33:32:43.7 (0.8) & $\text{PS260-N}$ & \texttt{point} & $51.0 \pm 15.2$ & $40.0 \pm 11.5$ & ~ \\  
        12:26:57.97 (0.50) &  +33:32:45.9 (0.6) & cluster tSZ & \texttt{circular Gaussian }& ~ & ~& $32.7 \pm 2.1$  \\ \hline
    \end{tabular}
    \tablefoot{ 
    Positional uncertainties are given in brackets.
    We indicate the function used to model each of the sources, the primary-beam-corrected flux of the point sources measured from the LSB ($\sim 74$~GHz) and USB ($\sim 90$~GHz) data, and the best-fitting size of the Gaussian function we used to describe the cluster signal (not corrected for the primary beam; see Sect.~\ref{sec:fitpress} for the fitted negative amplitude of the tSZ).}
    \label{tab:firstfit}
\end{table*}

\section{Data analysis}
\label{sec:dataanalysis}

The interferometric data were processed in various steps, with the goal of maximising the signal-to-noise ratio of the tSZ effect while carefully accounting for point sources in the field (see Fig.~\ref{fig:noemamaps}). We detail the construction of a source-clean continuum $uv$ table (using the \texttt{GILDAS} software\footnote{\url{https://www.iram.fr/IRAMFR/GILDAS/}}) and the detection of the tSZ effect in it below.

\subsection{Continuum: Combination of $uv$ tables}
\label{sec:continuum}

The spectral analysis of the data revealed two point sources in the field ($\text{PS260-N}$ and $\text{PS260-S}$; see Appendix~\ref{sec:pointsource}) with emission lines at $\sim 70.6$~GHz. To avoid contamination by the emission lines of the $\text{PS260-N}$ and $\text{PS260-S}$ regions, the data in the frequency
range below 70.78~GHz were removed from the spectral cube of the NOEMA visibilities in the further analyses.

\begin{figure}
    \centering
    \includegraphics[scale=0.42]{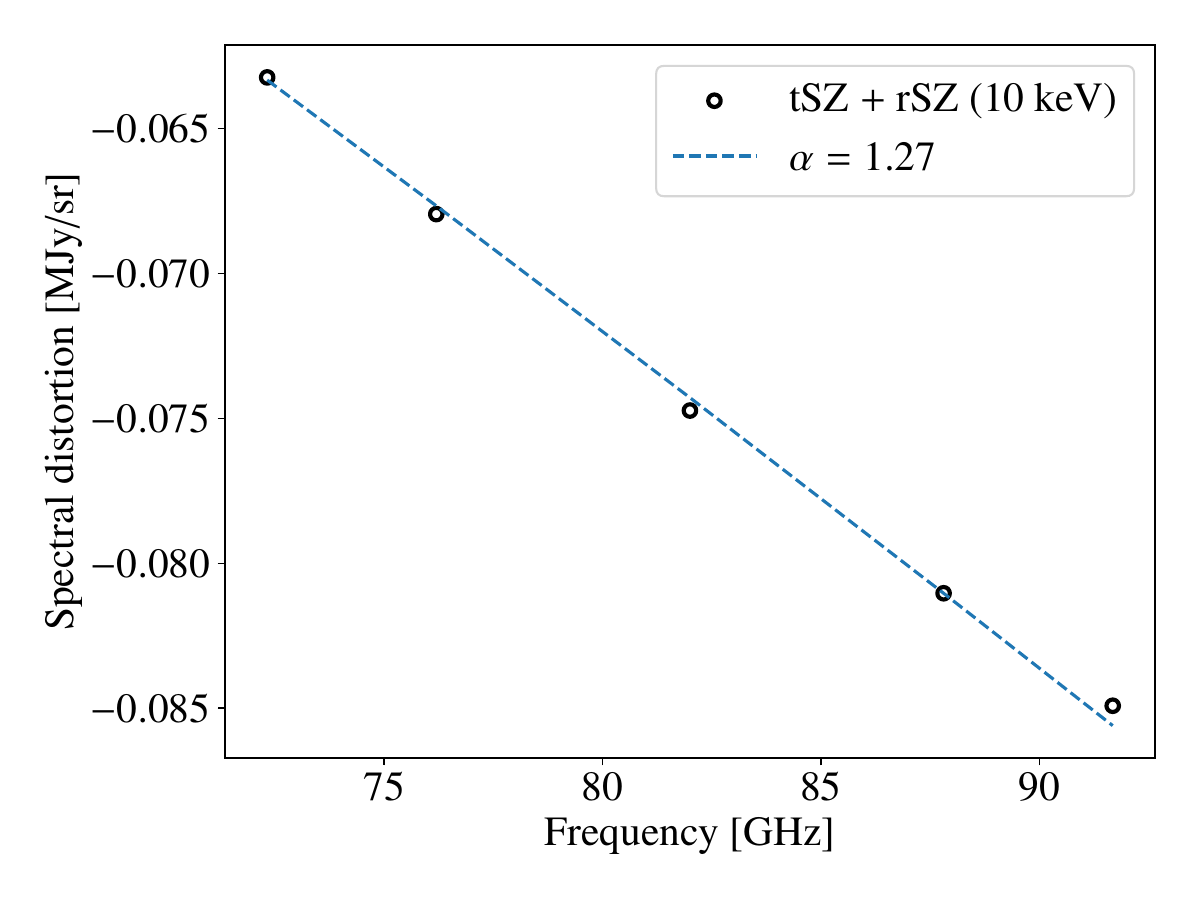}
    \caption{Spectral distortion of the CMB by the tSZ effect. The relativistic corrections assume 10 keV. We consider a Compton parameter of $y = 10^{-4}$. The black markers indicate the tSZ spectrum evaluated at the four central frequencies ($\nu_B = 72.320, 76.192, 87.808, 91.680$~GHz) and the mean frequency of the four basebands ($\nu_{82}=82$~GHz). The dashed line shows the power-law approximation we used to scale the visibilities.}
    \label{fig:tszdistortion}
\end{figure}

The line-free sections of the spectral cubes were combined to obtain a single continuum $uv$ table. First, four continuum $uv$ tables were computed by averaging the line-free frequency ranges of the polarisation-averaged basebands. This was done separately for the LSB and USB sidebands and resulted in two $uv$ tables for each sideband, at
72.320 and 76.192~GHz, and at 87.808 and 91.680~GHz central frequencies, respectively. 

To preserve the size and structure of the tSZ emission and maximise sensitivity, the $uv$ coordinates of the visibilities of the four $uv$ tables were 
scaled by a factor $\nu_B/\nu_{82}$, where $\nu_B$ and $\nu_{82}$ were the central frequencies ($\nu_B = 72.320, 76.192, 87.808, 91.680$~GHz) and the mean frequency of the four basebands ($\nu_{82}=82$~GHz), respectively. 
The real and imaginary parts of the visibilities of the four $uv$ tables were also corrected relative to the amplitude of the spectral CMB distortion by the factor $(\nu_B/\nu_{82})^{-\alpha}$, and the respective visibility weights were scaled by $(\nu_B/\nu_{82})^{2\alpha}$, where $\alpha$ is the spectral index of the distortion. We considered $\alpha = 1.27$ following the shape of the tSZ spectrum at the frequencies covered by the four basebands (Fig.~\ref{fig:tszdistortion}).
The four frequency-scaled and re-weighted $uv$ tables were concatenated into a single continuum $uv$ table. We accounted for the spectral shape of the tSZ emission and therefore optimised the data combination for the detection of the tSZ effect.

The concatenated $uv$ table was Fourier-transformed to generate a natural weighted dirty map and visually identify possible radio sources in the field that might contaminate the structure and signal of the tSZ effect. 
In addition to $\text{PS260-N}$ and $\text{PS260-S}$, a further radio source was identified near the core centre of \clj. 

The very central radio source detected by NOEMA at (RA, Dec)$_{\mathrm{J2000}}$ = (12$^{\rm h}$26$^{\rm m}$58.12$^{\rm s}$, +33$^\circ$32$'$48.7$^{\prime\prime}$) corresponds to the brightest cluster galaxy (BCG) in \clj\ \citep{2001ApJ...548L..23E,2009ApJ...693..617H, 2021NatAs...5..268D}, also known as PS9 \citep{2023A&A...671A..28M}. This galaxy was not detected by previous millimetre band observations, and given its central position in the cluster, it is crucial to consider its contamination. \citet{2023A&A...671A..28M} had to extrapolate the contribution of the BCG to the tSZ signal at 150~GHz from 1.4~GHz VLA FIRST Survey flux measurements \citep{1997ApJ...475..479W}. 

We removed the three point sources directly in the space of visibilities by making use of the \texttt{uv\_fit} routine included in the \texttt{MAPPING} software of \texttt{GILDAS}. First, the positions and fluxes of point sources were fitted together with the position, flux, and size of the CMB distortion on the concatenated $uv$ table. The fit was performed simultaneously for the point sources and the tSZ in order to minimise possible biases in the fitting of source positions. We modelled the three point sources with the \texttt{point} function in \texttt{uv\_fit}, which assumes that they are unresolved in the synthesized beam of NOEMA. The imprint of the tSZ effect on the CMB distortion was modelled with the \texttt{circular Gaussian} function of \texttt{uv\_fit} and assuming a negative amplitude. We discuss the impact that assuming a Gaussian shape for the cluster emission might have on the fitted point source fluxes in Sect.~\ref{sec:fitpress}.
Table~\ref{tab:firstfit} summarises the outputs of the fitting procedure on the concatenated $uv$ table. It is interesting to note that the PS9 flux measured from NOEMA data is consistent with the $3.60 \pm 0.13$~mJy measured by VLA FIRST at 1.4~GHz (Sect.~\ref{sec:continuum}) for a spectral index of $-0.97$, and this radio spectrum is coherent with the flux at 150~GHz inferred from the NIKA2 analysis by  \citet{2023A&A...671A..28M}.

Then, we considered the positions of the point sources measured in the concatenated table and the position and size of the tSZ to determine the associated fluxes of all the sources in each of the four initial $uv$ tables.
The spectral indices were subsequently fitted to the fluxes of the 
three point sources over the frequency range covered by the four $uv$ tables. The fluxes resulting from the spectral index measurements were removed at the fitted positions of the 
three point sources from each of the four $uv$ tables. Finally, we concatenated the four frequency-scaled re-weighted $uv$ tables after subtraction of the 
three point sources into one $uv$ table, which thus only contained the contribution from the tSZ effect.


\begin{figure}
    \centering
    \includegraphics[scale=0.455]{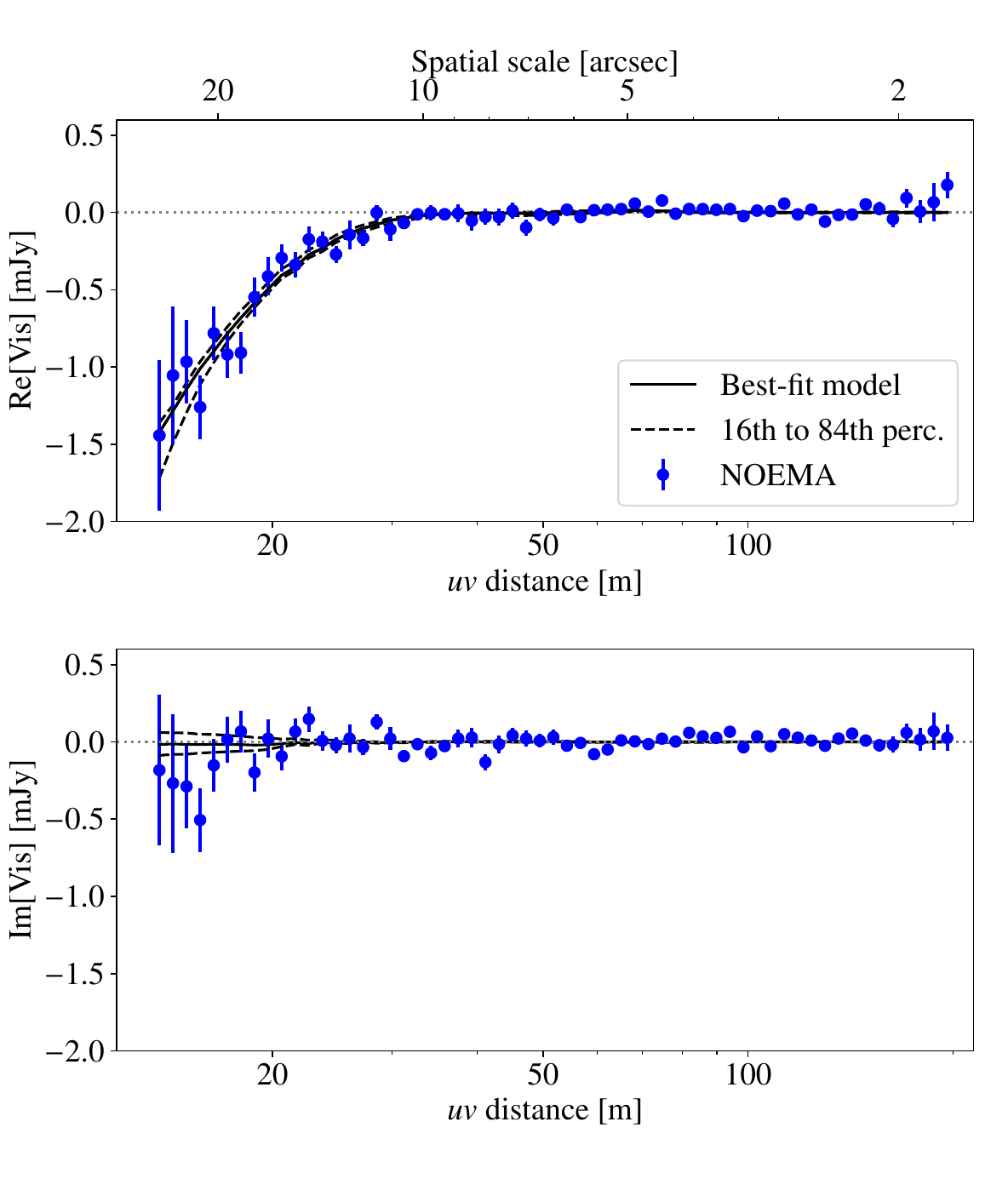}
    \caption{
    Real (top) and imaginary (bottom) visibility components as a function of binned $uv$ distances. The blue markers show the profile for the NOEMA data after subtraction of the point sources contribution from the visibilities. The error bars indicate $1\sigma$ uncertainties propagated from the error bar associated with each visibility. The solid black lines indicate the visibility profiles corresponding to the best-fitting model to the NOEMA data. The dashed black lines show the 16th to 84th percentiles of the fitted model.}
    \label{fig:tszraw}
\end{figure}
\begin{figure*}
    \centering
    \includegraphics[scale=0.226, trim={0cm 5cm 0cm 2cm} ]{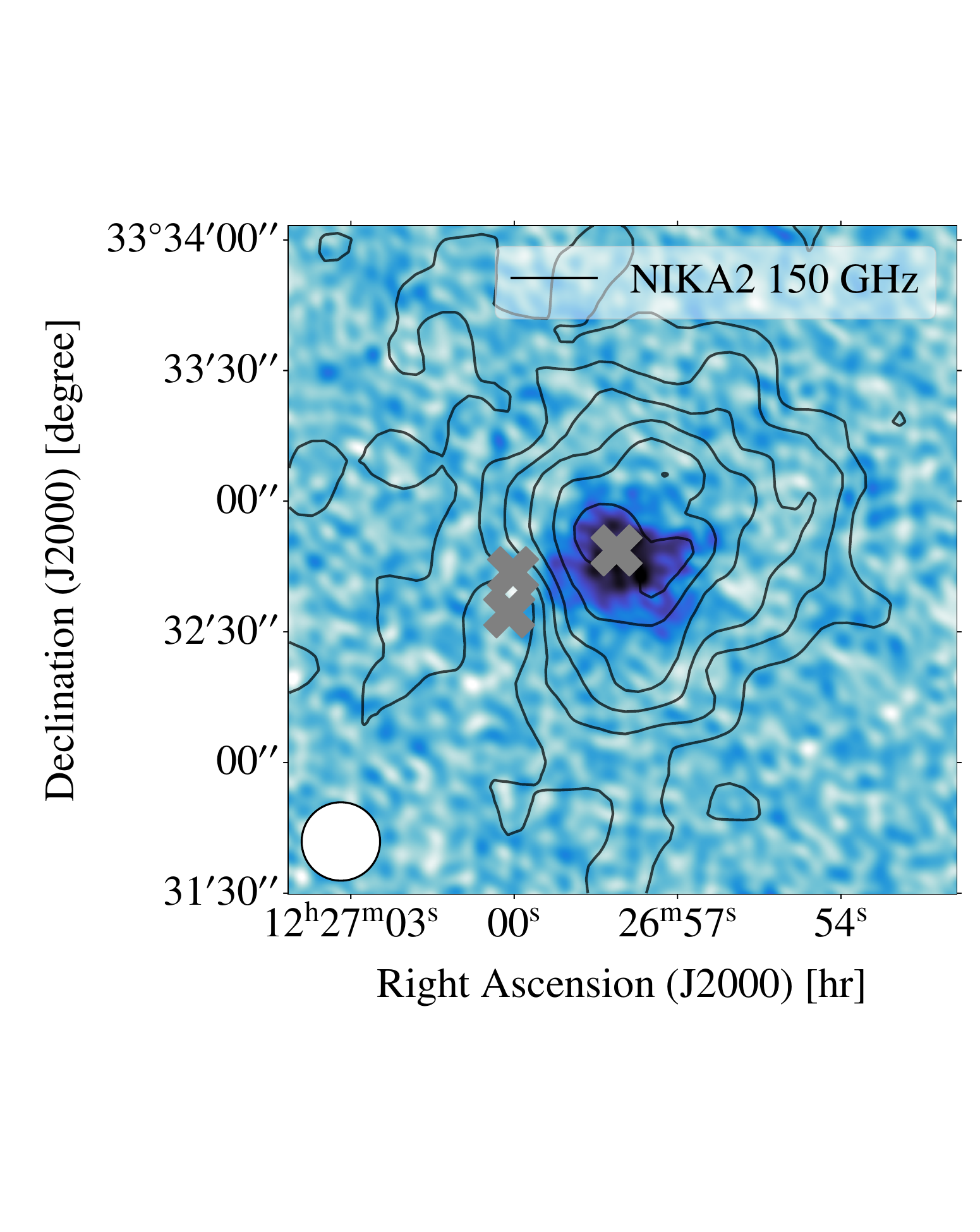}
    \includegraphics[scale=0.226, trim={0cm 5cm 0cm 2cm}]{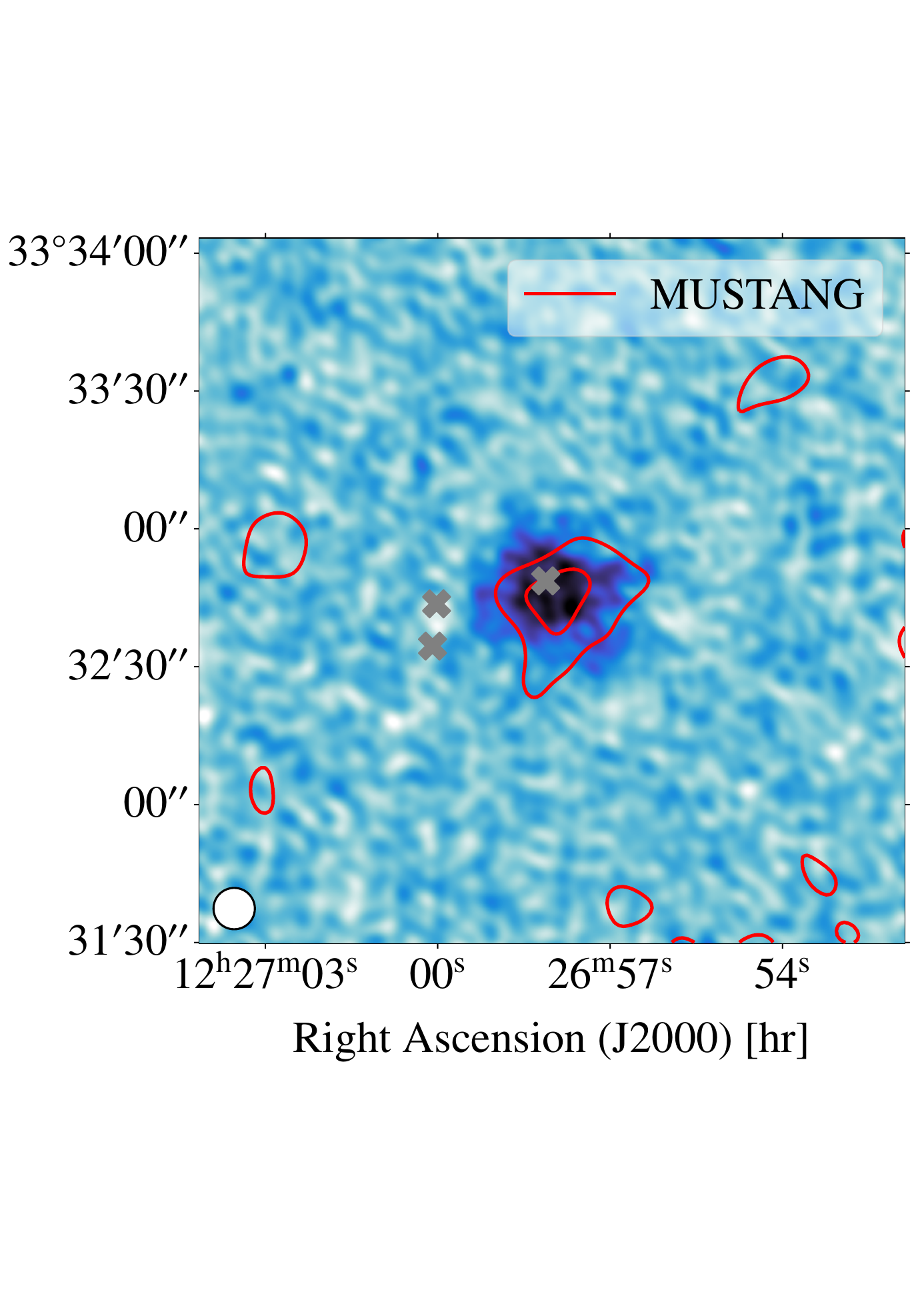}
    \includegraphics[scale=0.226, trim={0cm 5cm 0cm 2cm}]{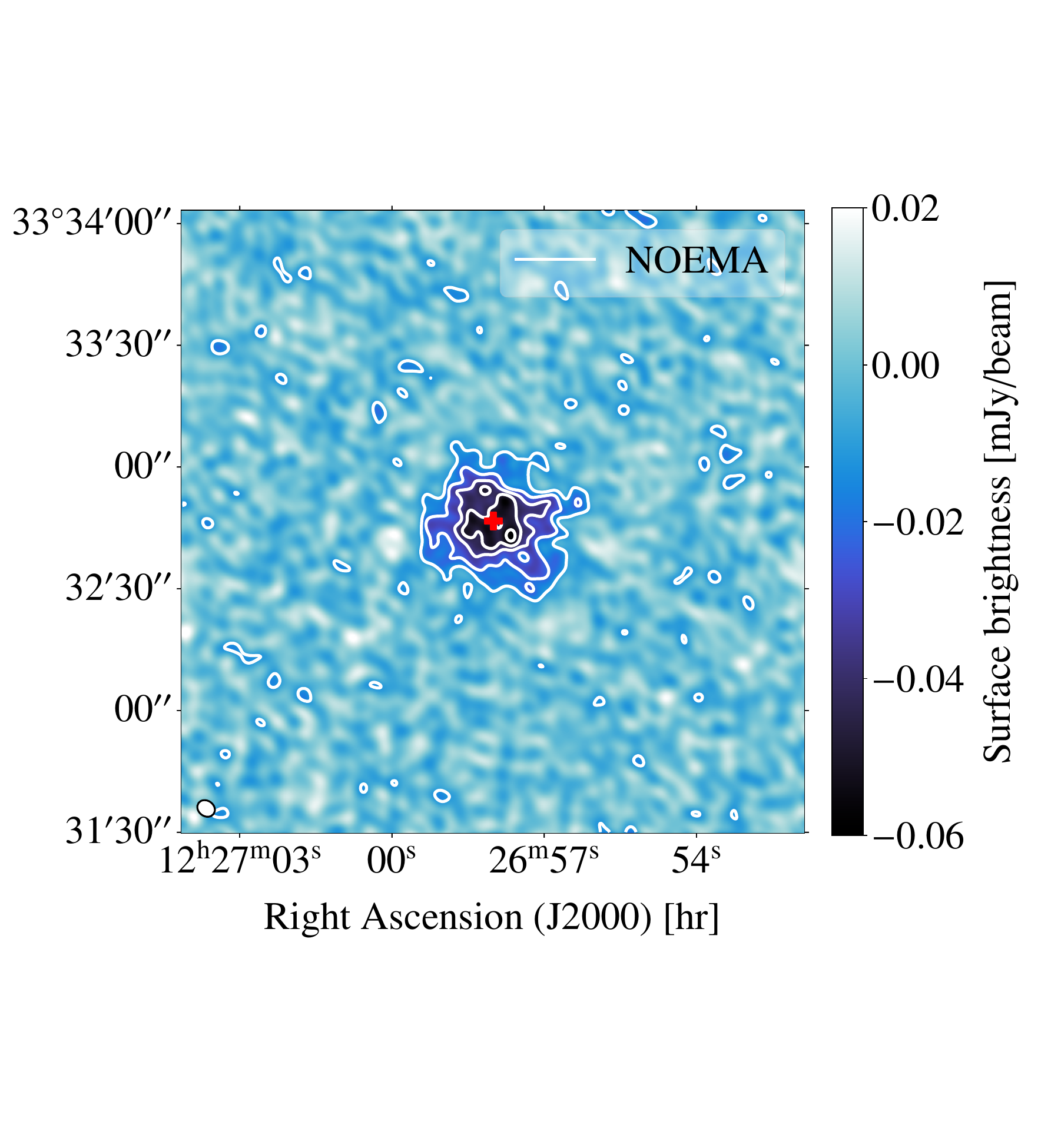}
    \caption{
    Point-source-subtracted and cleaned continuum NOEMA map of \clj\ compared to NIKA2 and MUSTANG observations.
    Left: NIKA2 150~GHz \citep{2023A&A...671A..28M} contours (black) starting from $2\sigma$ and spaced by $2\sigma$. The $18^{\prime\prime}$ beam is indicated in the lower left corner. Centre: MUSTANG 90~GHz \citep{2018A&A...612A..39R} contours (red) starting from $2\sigma$ and spaced by $2\sigma$ with the beam ($9^{\prime\prime}$) in the lower left corner. Right: $2\sigma$ levels in the NOEMA map (white). The size and orientation of NOEMA's synthesized beam (4.6$^{\prime\prime}\,\times\,$3.8$^{\prime\prime}$ at PA $=54^\circ$) are indicated in the lower left corner. The red cross indicates the centre of the map as defined from the tSZ peak according to NOEMA data (Table~\ref{tab:firstfit}). The grey crosses in left and central panels show the positions of the three sources we subtracted from the NOEMA data, but not from NIKA2 and MUSTANG to calculate contours.
    }
    \label{fig:tszmap}
\end{figure*}

\subsection{tSZ with NOEMA}
\label{sec:tsz}

\subsubsection{Visibility profiles}

By computing the average visibility values in concentric annuli from the point-source-free continuum $uv$ table (centred at the cluster position fitted in the previous section), we obtained the profiles shown with blue markers in Fig.~\ref{fig:tszraw}. The top panel presents the real part of the visibilities, and the bottom panel corresponds to the imaginary component. The negative signal in the real component of the visibilities is the evidence of the tSZ effect in our data. This is the first NOEMA detection of the tSZ effect on the CMB towards a galaxy cluster. The imaginary part also shows some negative signal, albeit less pronounced than the real component. We show in Sect.~\ref{sec:fitpress} that this is related to a departure from circular symmetry of the data.

\subsubsection{Continuum map}

The visibility measurements 
were Fourier-transformed to generate a natural weighted dirty map of the tSZ effect CMB distortion using the \texttt{GILDAS} software package. The map was cleaned with the \texttt{CLARK} algorithm \citep{1980A&A....89..377C} down to a $1\sigma$ noise level of 5.6~$\mu$Jy beam$^{-1}$ and is shown in the bottom panel in Fig.~\ref{fig:noemamaps}. 
The array configuration and $uv$ coverage, using natural weighting, yielded a synthesized beam of 4.8$^{\prime\prime}\,\times\,3.6^{\prime\prime}$ with a position angle (PA) of 54$^\circ$ at the frequency of 82\,GHz (shown in the lower left corner in Fig.~\ref{fig:noemamaps}). The negative extended emission at the centre of the map corresponds to the tSZ effect in \clj.

For comparison, we show in Fig.~\ref{fig:tszmap} the contours corresponding to MUSTANG 90~GHz \citep[centre, ][]{2018A&A...612A..39R} and NIKA2 150~GHz \citep[left, ][]{2023A&A...671A..28M} observations of \clj. NOEMA provides an unprecedentedly highly resolved image of the cluster core.

\section{Pressure profile of \clj }
\label{sec:pressure}

\subsection{Fit of a pressure model to NOEMA visibilities}
\label{sec:fitpress}

\begin{table}[]
    \caption{Prior distributions for the parameters in the model we fitted to NOEMA visibilities.}
    \centering
    \begin{tabular}{c c}
     \hline  \hline
        Parameter & Prior distribution  \\\hline
        $P_0,P_1,P_2,P_3 $ [keV/cm$^3$]  & $\mathcal{U}(0, 2)$   \\
        Res. PS9, PS260-S, PS260-N [Jy]  & $\mathcal{U}(-1, 1)$   \\ 
        $\Delta$RA$_{\mathrm{J2000}}$ [arcmin]  & $\mathcal{U}(-1, 1)$   \\ 
        $\Delta$Dec$_{\mathrm{J2000}}$ [arcmin]  & $\mathcal{U}(-1, 1)$  \\ 
        Calib [MJy/sr]  & $\mathcal{N} (-747.30, 77.58^2 )$   \\ 
        \hline
    \end{tabular}

    \label{tab:freeparams_firstfit}
\end{table}
\begin{figure*}
    \sidecaption
    \includegraphics[width=12cm, trim={2cm 0.7cm 2cm 1cm}]{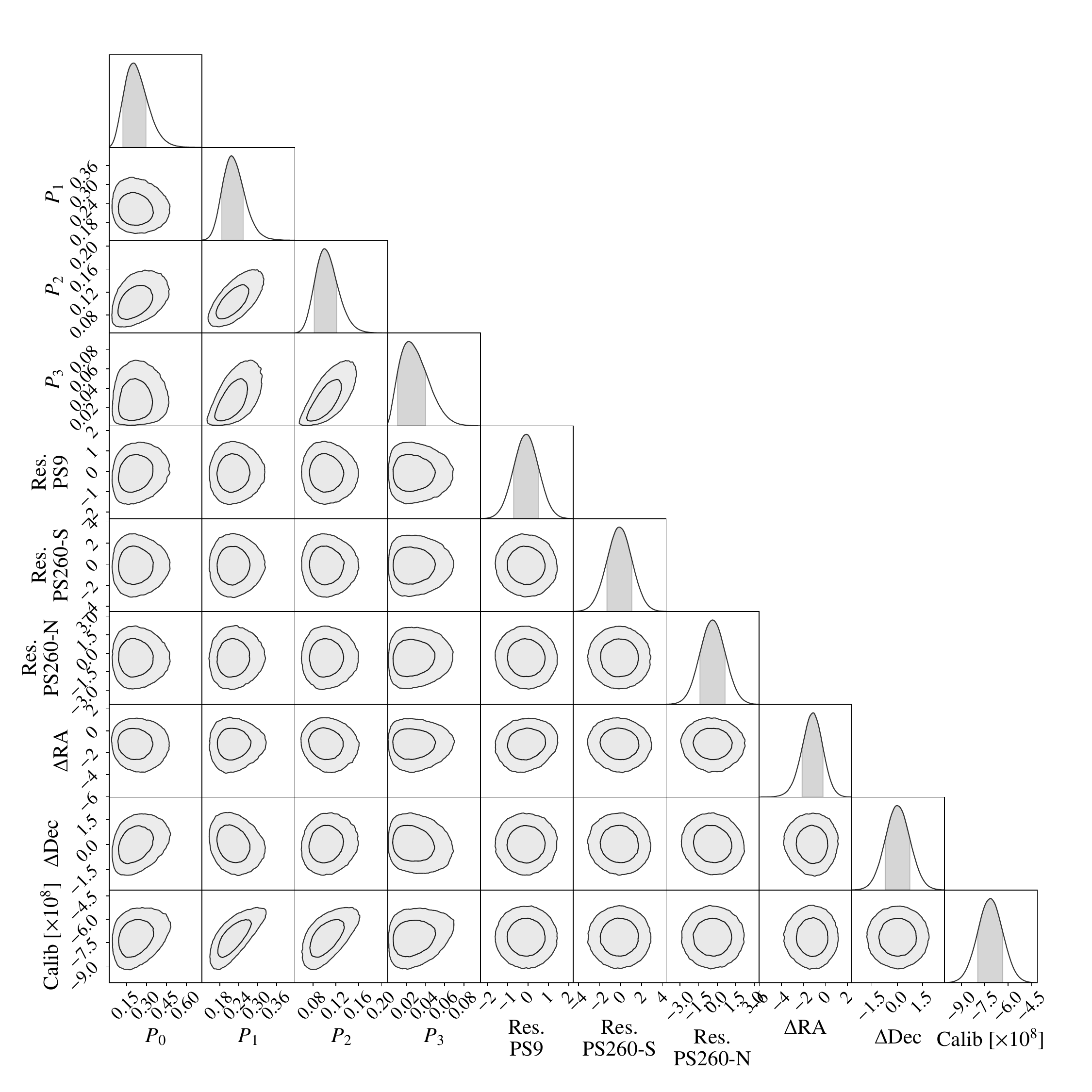}
    \caption{Posterior probability distributions (1D and 2D) of the free parameters in the model fitted to the NOEMA visibilities. The pressure bins ($P_0, P_1, P_2,$ and $P_3$) are defined at 15, 60, 150, and 270~kpc and given here in keV/cm$^3$ units. The residual point source fluxes (Res. PS9, Res. $\text{PS260-S}$, and Res. $\text{PS260-N}$) are in $10^{-5}$ Jy units, $\Delta$RA$_{\mathrm{J2000}}$ and $\Delta$Dec$_{\mathrm{J2000}}$ offsets are shown in arcseconds, and the calibration factor (Calib) is in Jy/sr.
    }
    \label{fig:corner}
\end{figure*}

\begin{figure}
    \centering
    \includegraphics[scale=0.362, trim={1cm 0.cm 0cm 0cm}]{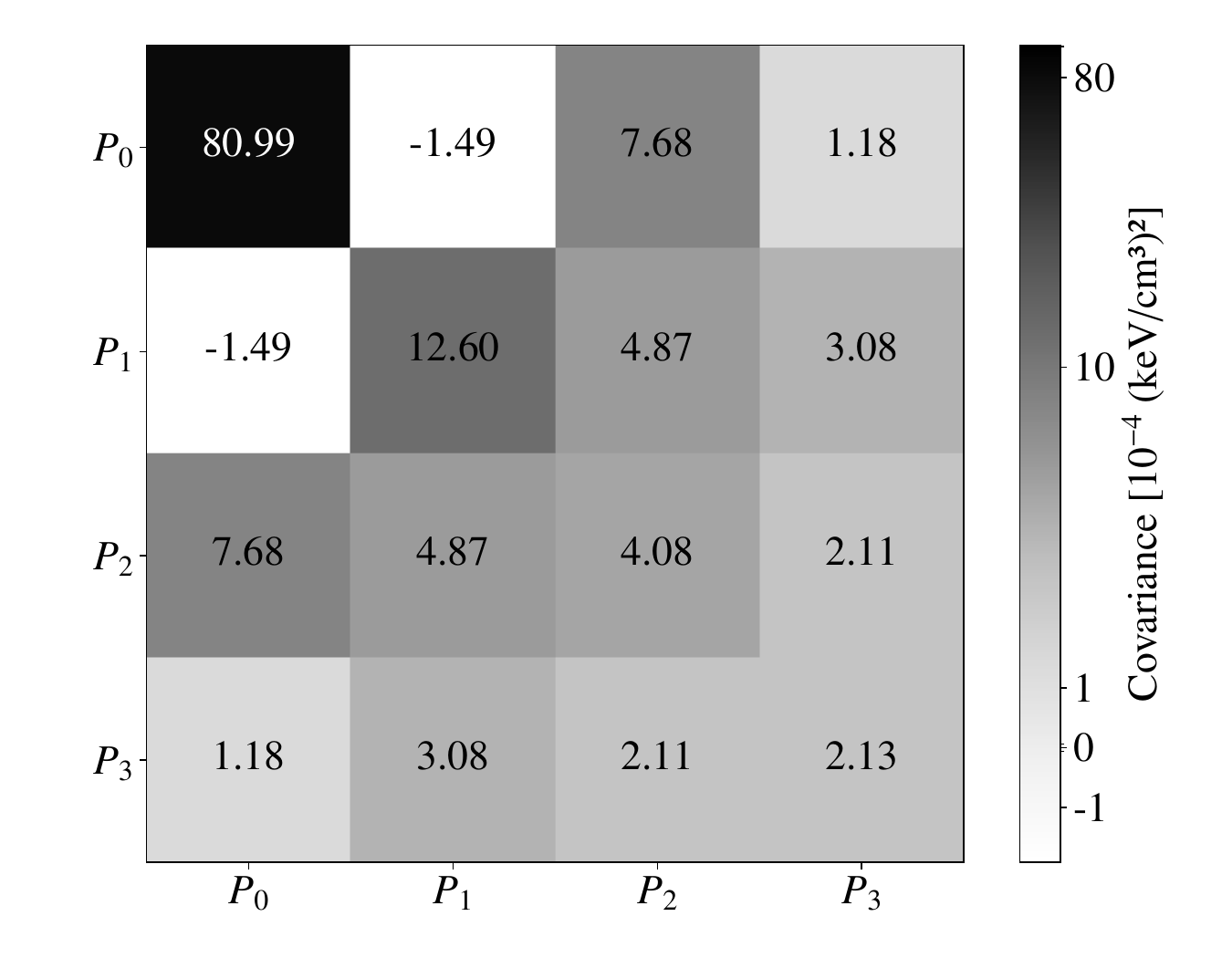}
    \caption{Covariance matrix of the pressure bins we fitted to NOEMA visibilities in units of $10^{-4}$ (keV/cm$^{3}$)$^{2}$.}
    \label{fig:covmat_bins}
\end{figure}

The thermal Sunyaev-Zel'dovich effect traces the thermal pressure of the electron gas ($P_{e}$) in the ICM of clusters \citep{1980ARA&A..18..537S}. Thus, the high angular resolution tSZ observations by NOEMA can be used to extract valuable constraints on the thermal pressure distribution in the cluster cores.

We modelled the visibilities measured by NOEMA for \clj\ with a spherical pressure profile. We described the electron pressure, $P_{e}$, with a power-law model \citep[also known as binned or non-parametric,][]{2010A&A...519A..29B, 2017A&A...597A.110R,2018A&A...612A..39R, 2018A&A...615A.112R, 2023A&A...671A..28M, 2023OJAp....6E...9K, 2024A&A...684A..18A},
\begin{equation}
    P_{e}( r_{i} <r<r_{i+1}) = P_i \left( \frac{r}{r_{i}} \right)^{-\alpha_{i}},
\end{equation}
with
\begin{equation}
    \alpha_i = - \frac{\mathrm{log} P_{i+1} - \mathrm{log}P_i}{\mathrm{log}r_{i+1}  - \mathrm{log} r_i}.
\end{equation}

The free parameters of the model are the values of the pressure ($P_i$) at fixed radii from the centre of the cluster ($r_i$). In particular, we fitted the pressure at 15, 60, 150, and 270~kpc, spanning the radial range probed by these NOEMA data. We verified that fitting more bins introduces correlations amongst them and does not improve the constraints on the pressure.

At each step of the fit, the 
three-dimensional pressure profile is integrated along the line of sight to obtain a Compton map. We used the software \texttt{minot}\footnote{\url{https://github.com/remi-adam/minot}} \citep{2020A&A...644A..70A} to compute the model.
Possible differences between the assumed cluster centre (Table~\ref{tab:firstfit}) and the best-fitting position were accounted for by introducing offsets in Right Ascension ($\Delta$RA$_{\mathrm{J2000}}$) and Declination ($\Delta$Dec$_{\mathrm{J2000}}$).
The Compton map was converted into a tSZ surface brightness map by multiplying it with a calibration factor. We considered this calibration factor as a free parameter and took a Gaussian prior: Calib $\sim \mathcal{N} (-747.30, 77.58^2 )$ MJy/sr. The central value corresponds to the calibration factor of the tSZ signal at 82~GHz when the relativistic effect in a 10~keV intracluster medium is accounted for \citep{2007ApJ...659.1125M}. The $\sim 10 \%$ scatter is given by the combination of the observation calibration accuracy (Sect. \ref{sec:target}) and the relative difference between the tSZ calibration factor for a 10~keV 
and a 0~keV cluster. These are the main error sources in the calibration.

We measured the point source fluxes and removed them from the visibilities by assuming a two-dimensional circular Gaussian distribution for the tSZ emission of the cluster (Sect.~\ref{sec:continuum}). The shape of the tSZ signal from clusters is known to be different from a Gaussian (Sect.~\ref{sec:resolvedpress}). This approximation for the tSZ signal might therefore have led to biased estimates of point source fluxes, and consequently, to a residual presence of point sources in the source-clean data. To account for a possible residual contamination, we also modelled the three sources by fitting their fluxes considering that they are unresolved sources located at the positions indicated in Table~\ref{tab:firstfit}. Finally, we multiplied the model map, including both the tSZ model and the three point sources, by the effective primary beam\footnote{We verify in Appendix~\ref{sec:simulationfit} that assuming this effective attenuation, rather than a frequency-dependent primary beam attenuation, does not bias the reconstructed pressure.} (56.2$^{\prime \prime}$ FWHM) to account for the attenuation of the sky signal introduced by the response of each antenna in the observations. Table~\ref{tab:freeparams_firstfit} summarises the free parameters we considered in the fit and the priors on each.

Raw interferometric data and the associated uncertainties being defined in the $uv$ plane, it is preferred to model the signal in the visibility space instead of in the image space. Thus, we converted the model map into visibilities that can be directly compared and fitted to the measured ones. For this purpose, we called at each step of the fit the package \texttt{galario}\footnote{\url{https://mtazzari.github.io/galario/index.html}} \citep{2018MNRAS.476.4527T}. 
The attenuated model map was given to \texttt{galario} together with the table containing the visibilities, the corresponding weights $W_i$ (inverse of the variance related to each visibility), and their positions in the $uv$ plane for the NOEMA data. As an output, \texttt{galario} provided the input model map in the visibility space. The output model sampled the same $u$ and $v$ positions as the NOEMA data. The log-likelihood is given by
\begin{equation}
    \ln \mathcal{L} = - \frac{1}{2} \sum_i \left[ (\Delta R_i^2 + \Delta I_i^2) W_i \right],
\end{equation}
where $\Delta R_i$ and $\Delta I_i$ correspond to the difference between data and model visibilities for the real and imaginary components at each $uv$ position $i$.

The fit was performed using a Markov chain Monte Carlo (MCMC) algorithm with the software \texttt{emcee}  \citep{2019JOSS....4.1864F, 2010CAMCS...5...65G}. We used 90 walkers and $5 \times 10^4$ steps. First $2 \times 10^4$ steps were discarded, and convergence was confirmed following the $\hat{R}$ test \citep{gelmanrubin} and autocorrelation of chains. The robustness of the fitting procedure was tested on simulations (see Appendix~\ref{sec:simulationfit}). We show in Fig.~\ref{fig:corner} the one- and two-dimensional posterior distributions of the fitted parameters (Table~\ref{tab:freeparams_firstfit}), that is, the values of the pressure at the mentioned radial bins, the residual fluxes of the point sources, the offsets with respect to the assumed cluster centre, and the Compton-tSZ calibration factor. The residual fluxes of all three point sources are compatible with zero, and the best-fitting tSZ centre has an offset of ($\Delta$RA, $\Delta$Dec)$_{\mathrm{J2000}}$ = ($-1.32, 0.04$) arcsec with respect to the centre measured directly in the NOEMA data (Table~\ref{tab:firstfit}).

\renewcommand{\arraystretch}{1.4}   
\begin{table}[]
\caption{Best-fit values and error bars at the 16th and 84th percentiles for the pressure model fitted to NOEMA visibilities.}
    \centering
    \begin{tabular}{c c}
    \hline\hline
        Radius  &  Pressure \\ 
         $[$kpc$]$   &  [keV/cm$^3$] \\ \hline
        15     &   $0.163^{+0.153}_{-0.023}$ \\ 
        60     &  $0.210^{+0.050}_{-0.019}$ \\
        150    &   $0.090^{+0.035}_{-0.005}$ \\
        270    &   $0.019^{+0.025}_{-0.004}$ \\ \hline
    \end{tabular}
    \label{tab:fitbestfit}
\end{table}

As aforementioned, in Fig.~\ref{fig:tszraw} we show the real and imaginary visibility profiles obtained by radially averaging the NOEMA visibilities in the $uv$ coordinates. We also present the visibility profile that corresponds to the best-fitting model together with the 16th and 84th percentiles. The model and data visibility profiles were centred at the tSZ position that was initially measured in the NOEMA data (Table~\ref{tab:firstfit}), with the cluster tSZ signal in the model map centred at the fitted position (Fig.~\ref{fig:corner}). Thus, the negative imaginary part in the data (Fig.~\ref{fig:tszraw}) reflects asymmetric structures that might be due to substructures or disturbances in the intracluster medium that are not reproduced by the fitted spherical pressure model.

We summarise the 
best-fit values for the pressure bins together with 16th and 84th percentiles 
calculated from the marginalised posterior probability distributions in Table~\ref{tab:fitbestfit}. Their covariance matrix is given in Fig.~\ref{fig:covmat_bins}. Black empty diamonds in Fig.~\ref{fig:bestfitgnfw} show the best-fit three-dimensional pressure profile as a function of physical radii. In the same figure, we compare the NOEMA fit to other pressure reconstructions obtained from millimetric observations of \clj\ with NIKA, NIKA2, MUSTANG, and Bolocam instruments installed at single dish telescopes \citep{2015A&A...576A..12A, 2018A&A...612A..39R, 2023A&A...671A..28M}. The agreement of NOEMA with previous results is very good within the common radial range, and this allowed us to combine the different datasets to obtain improved constraints on the pressure profile of \clj. NOEMA has provided a new piece of information here with the measurement of the pressure at the cluster core.

\subsection{Resolved pressure profile of \clj }
\label{sec:resolvedpress}

\begin{figure}
    \centering
    \includegraphics[scale=0.355, trim={0.5cm 0 0 0}]{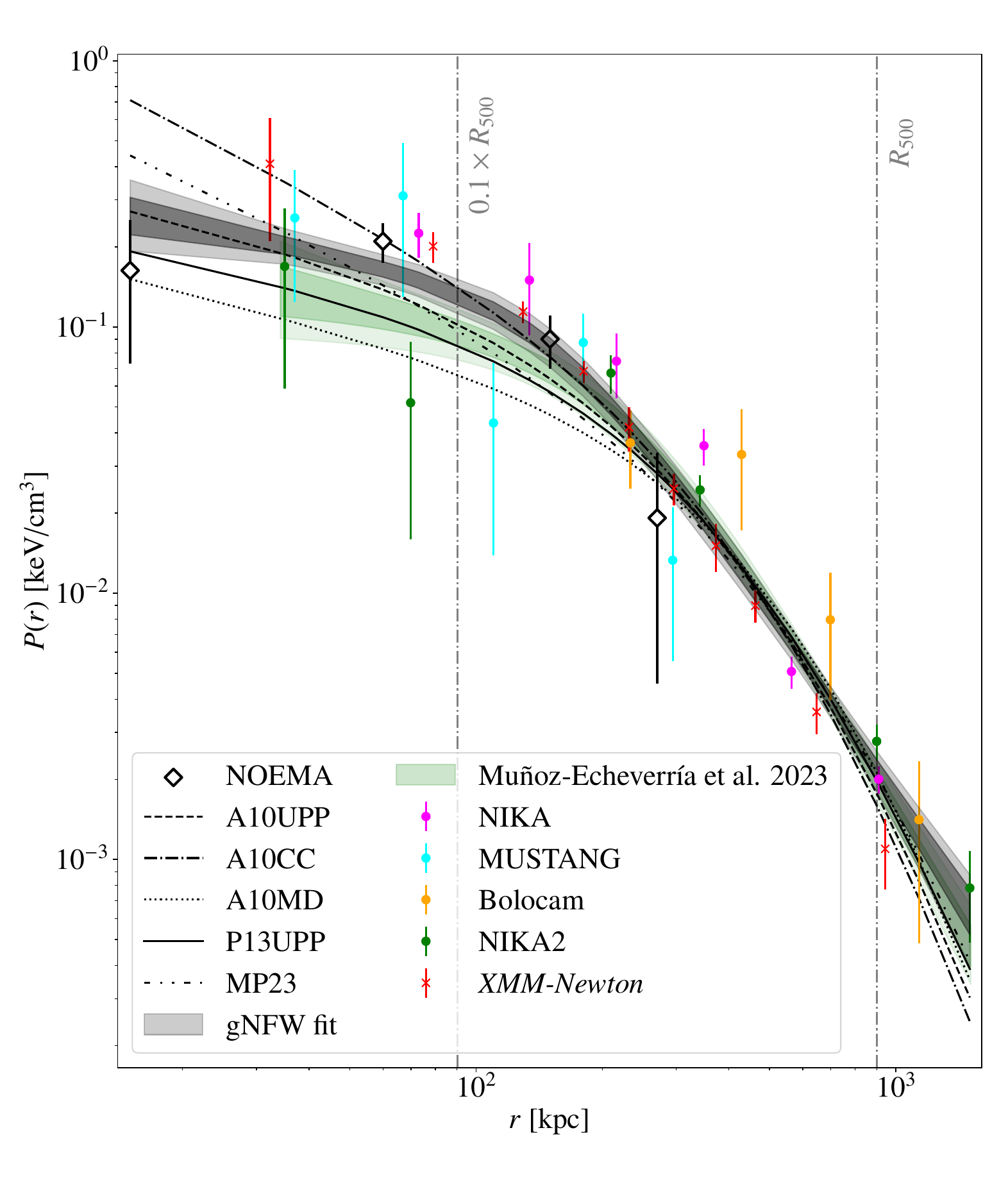}
    \caption{
    Pressure profile of \clj. The empty black diamonds indicate the best pressure profile bins fitted to NOEMA visibilities. The magenta, blue, and orange markers show the profiles reconstructed by \citet{2018A&A...612A..39R} for NIKA, MUSTANG, and Bolocam data, respectively. The green markers correspond to the NIKA2 pressure reconstruction from \citet{2023A&A...671A..28M}. The pressure profile reconstructed from \textit{XMM-Newton} X-ray observations \citep{2023A&A...671A..28M} is shown in red. The error bars represent
    the square root of the diagonal elements of the covariance matrix.
    The grey shaded areas indicate the 16th to 84th and 2.5th to 97.5th percentiles for the gNFW pressure profile we fitted to all tSZ pressure bins. In green, we present the gNFW profile obtained by \citet{2023A&A...671A..28M} without NOEMA data and neglecting the correlation of the pressure bins from \citet{2018A&A...612A..39R}. The dashed, dash-dotted, and dotted lines correspond to the universal, cool-core, and morphologically disturbed pressure profiles from \citet{2010A&A...517A..92A}, respectively. The solid and dash-dot-dotted profiles show the pressure profiles obtained by \citet{2013A&A...550A.131P} and \cite{2023A&A...678A.197M}, respectively. The vertical dashed-dotted grey lines indicate the $0.1 \times R_{500} $ and $R_{500}$ characteristic radii of the cluster.
    }
    \label{fig:bestfitgnfw}
\end{figure}

\begin{figure}
    \centering
    \includegraphics[scale=0.42, trim={0.5cm 0.2cm 0cm 0cm}]{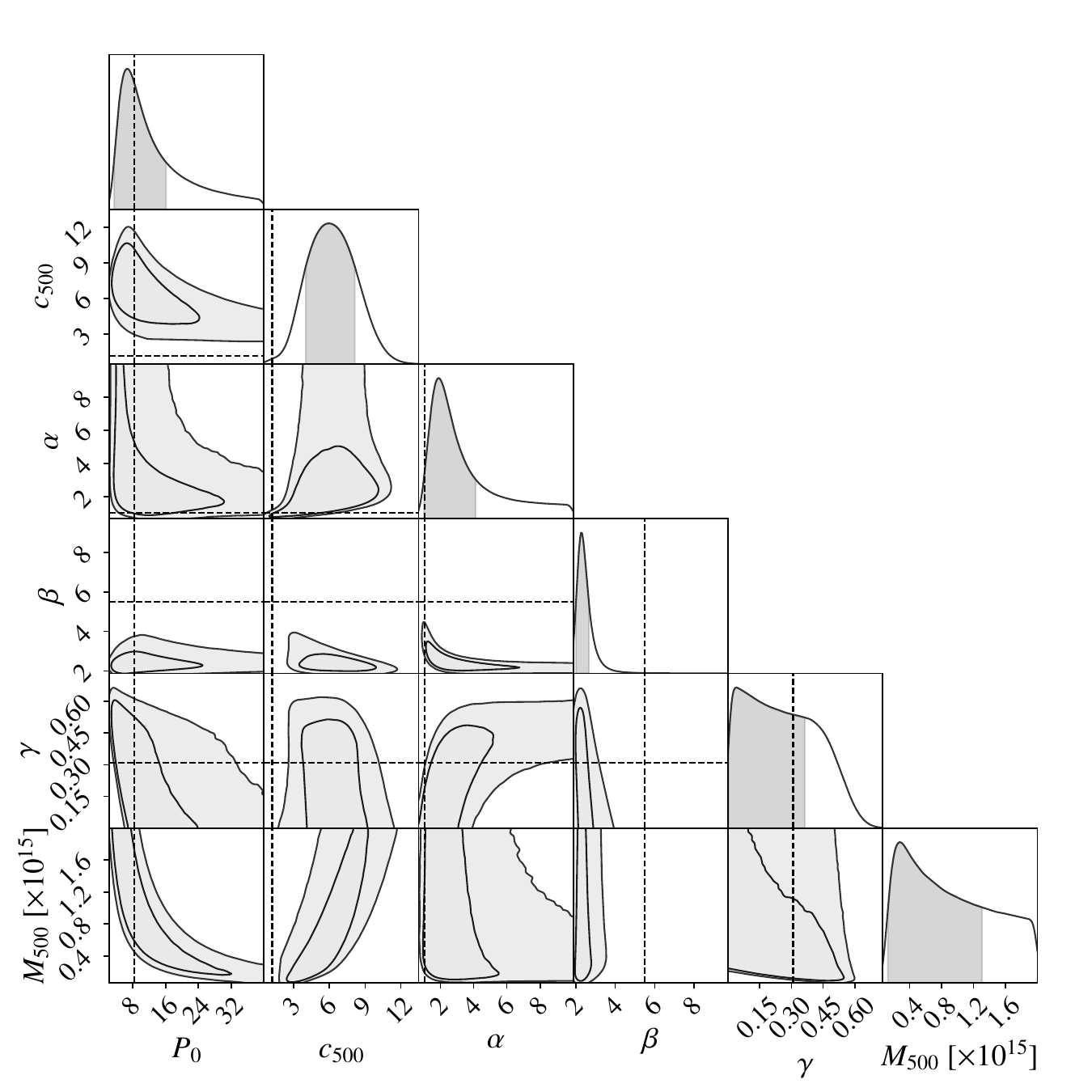}
    \caption{Posterior probability distributions of the gNFW model parameters jointly fitted to individual NOEMA, NIKA2, NIKA, MUSTANG, and Bolocam pressure reconstructions. The dashed lines indicate the parameter values obtained by \citet{2010A&A...517A..92A}. We also show the posterior distribution for the mass in solar mass units.}
    \label{fig:gnfwparams}
\end{figure}

Following the approach considered in previous works \citep{2018A&A...612A..39R, 2018A&A...615A.112R, 2023A&A...671A..28M}, we combined the binned pressure profiles reconstructed from different instruments to jointly fit a global pressure model for \clj. We parametrised the pressure with a generalised Navarro-Frenk-White model \citep[gNFW,][]{2007ApJ...668....1N} and searched to minimise

\begin{equation}
\label{eqn:chi2}
    \begin{split}
    \chi^2 & = \sum_{\mathrm{instr}}  \chi^2_{\mathrm{instr}} \\ 
    &=  \sum_{\mathrm{instr}}  \left( P_{\mathrm{gNFW}}(\vartheta) - P_{\mathrm{instr}}\right)^{T} C_{\mathrm{instr}}^{-1} \left(P_{\mathrm{gNFW}}(\vartheta) - P_{\mathrm{instr}}\right). \\    
     \end{split}     
\end{equation}

Here, $P_{\mathrm{instr}}$ and $C_{\mathrm{instr}}$ are the pressure bins and their associated covariance matrix reconstructed from the tSZ map of a given instrument, with ``instr''~=~[NIKA, MUSTANG, Bolocam, NIKA2, NOEMA]. $P_{\mathrm{NOEMA}}$ and $C_{\mathrm{NOEMA}}$ were obtained from the fit to the NOEMA visibilities in Sect.~\ref{sec:fitpress}. $P_{\mathrm{NIKA2}}$ and $C_{\mathrm{NIKA2}}$ correspond to the pressure bins (green markers in Fig.~\ref{fig:bestfitgnfw}) and covariance matrix obtained with the ``Time Ordered 2D'' approach from NIKA2 data by \citet{2023A&A...671A..28M}. In addition,
$P_{\mathrm{NIKA}}$, $C_{\mathrm{NIKA}}$, $P_{\mathrm{MUSTANG}}$, $C_{\mathrm{MUSTANG}}$, $P_{\mathrm{Bolocam}}$, and $C_{\mathrm{Bolocam}}$ are the pressure values and associated covariances reconstructed by \citet{2018A&A...612A..39R}\footnote{Pressure profiles in \citet{2018A&A...612A..39R} and in this work are centred on positions at 3 arcsec of distance, which is within the angular resolution of NOEMA data given by the synthesized beam.} from the analysis of NIKA, MUSTANG, and Bolocam maps, respectively. These values are given in Table 2 in \citet{2018A&A...612A..39R} and are shown with magenta, blue, and orange markers in Fig.~\ref{fig:bestfitgnfw}. 
    
Finally, $P_{\mathrm{gNFW}}(\vartheta)$ in Eq.~\ref{eqn:chi2} corresponds to the gNFW pressure model for a set of parameters $\vartheta=[P_{0}, c_{500}, \alpha, \beta, \gamma]$. At each physical radius $r$, the model is defined as
\begin{equation}
\label{eqn:gnfw}
    P_{\mathrm{gNFW}}(r) =  P_{500} \frac{P_0 }{(c_{500}r/R_{500})^\gamma \left[ 1 + (c_{500}r/R_{500})^{\alpha}\right]^{(\beta-\gamma)/\alpha} },
\end{equation}
with $R_{500}$ the radius of \clj\ at which the mean mass density of the cluster is 500 times the critical density of the Universe at its redshift. Here, $P_{500}$ is also a function of the mass and redshift of the cluster that we computed following Eqs.~5 and 13 in \cite{2010A&A...517A..92A}. We propagated the uncertainty on the $M_{500}$ mass of \clj\ by including the mass as a free parameter in the fit. We considered a Gaussian prior distribution based on the $M_{500} = 5.70^{+0.63}_{-0.69} \times 10^{14} \hspace{2pt}\mathrm{M}_{\odot}$ estimate from \citet{planck2016} (Table~\ref{tab:freeparams_gnfwfit}). At each step of the fit, we calculated the corresponding $R_{500}$ and $P_{500}$ to compute the pressure profile model (Eq.~\ref{eqn:gnfw}).

\begin{table}[]
\caption{Prior distributions for the parameters in the gNFW model we fitted to the pressure bins. }
    \centering
    \begin{tabular}{c c}
    \hline \hline
        Parameter & Prior distribution  \\ \hline
        $P_0$ & $\mathcal{U}(0, 40)$ \\ 
        $c_{500}$ & $\mathcal{U}(0, 20)$ \\
        $\alpha$ & $\mathcal{U}(0, 10)$ \\
        $\beta$ & $\mathcal{U}(0, 20)$ \\
        $\gamma$ & $\mathcal{U}(0, 10)$ \\
        $M_{500}$ [$10^{14}$ M$_{\odot}$] & $\mathcal{N}(5.70, 0.66^2)$ \\ \hline
    \end{tabular}
    \tablefoot{For the $M_{500}$ mass parameter, we took a Gaussian prior based on the estimate from \citet{planck2016}, truncated at $20 \times 10^{14}$ M$_{\odot}$.}
    \label{tab:freeparams_gnfwfit}
\end{table}

We performed an MCMC fit again using the \texttt{emcee} software and the above-mentioned convergence criteria. The prior distributions for the fitted parameters are given in Table~\ref{tab:freeparams_gnfwfit}, and their posterior probability distributions are shown in Fig.~\ref{fig:gnfwparams}. Dashed lines in Fig.~\ref{fig:gnfwparams} indicate the gNFW profile parameter values obtained in \citet{2010A&A...517A..92A}.
The corresponding pressure profile is shown in grey in Fig.~\ref{fig:bestfitgnfw}, where shaded areas represent 16th to 84th and 2.5th to 97.5th percentiles. The vertical dashed-dotted grey lines indicate $R_{500}$ and $0.1 \times R_{500}$, the characteristic radius beyond which clusters are expected to be self-similar \citep{2010A&A...517A..92A,2023ApJ...944..221S}.
For comparison, we present in Fig.~\ref{fig:bestfitgnfw} the gNFW profile reconstructed by \citet{2023A&A...671A..28M} without NOEMA data. \citet{2023A&A...671A..28M} did not consider the covariances of NIKA, MUSTANG, and Bolocam pressure bins either. It is thanks to the
NOEMA observations that we have been able to constrain the thermal pressure content in the core of \clj. 
We precisely measured the tSZ effect below 40~kpc for the first time in this cluster 
and placed tight constraints on the pressure profile at $\sim$100~kpc. As a result, we can now probe the distribution of the thermal gas in \clj\ in a radial range spanning two orders of magnitude.

In Fig.~\ref{fig:bestfitgnfw}, we also present in dashed, dash-dotted, and dotted lines the "Universal" (A10UPP), "Cool-Core" (A10CC), and "Morphologically-Disturbed" (A10MD) pressure profiles reconstructed in \citet{2010A&A...517A..92A}. The solid profile corresponds to the radial pressure distribution as estimated from the analysis of \textit{Planck} and \textit{XMM-Newton} data in \citet{2013A&A...550A.131P}, P13UPP, and the dash-dot-dotted profile shows the best gNFW model obtained from the combination of South Pole Telescope (SPT) and \textit{Planck} data in \cite{2023A&A...678A.197M}, MP23. The slope of the pressure in the core of \clj\ is very flat, which is indicative of a non-relaxed dynamical state \citep{2010A&A...517A..92A}. This agrees with other indications \citep{2009ApJ...691.1337J,2011ApJ...734...10K} for a disturbed system. In particular, the flattening of the thermal pressure profile might be due to the ICM sloshing \citep[see][and references therein]{2019ApJ...882..119Z,2020A&A...633A..42S} from the interaction of the two dark matter clumps that form the cluster \citep[as concluded from the weak lensing analysis in][]{2009ApJ...691.1337J}. \cite{2018A&A...614A.118A} reported indications of discontinuities towards the south-west of the core in the NIKA tSZ map (see also the NIKA2 contours in the left panel in Fig.~\ref{fig:tszmap}), which agree with the signature expected for a shock and in the direction of the temperature excess measured from X-rays by \cite{2007ApJ...659.1125M}. This scenario might introduce cold fronts and non-thermal pressure contributions, even maybe turbulence, and would be supported by the diffuse radio emission observed by \citet{2021NatAs...5..268D}. The deviation from spherical symmetry of the gas distribution that we observed with NOEMA (Sect.~\ref{sec:fitpress}) also agrees with an unrelaxed dynamical state. Further analyses of NOEMA data combined with single-dish maps of \clj\ could include an elliptical modelling of the cluster, and thus, fully exploit the power of interferometry to probe structure in this distant object.

The flattening at galaxy scales ($15$~kpc) that we measured might also originate from the interplay between the ICM and the circumgalactic medium \citep[CGM,][]{2017ARA&A..55..389T} of the central galaxy in \clj. This result might indicate that the gas in the ICM has cooled down in the very core of the cluster, in the ICM-CGM interface, without a stark drop in density \citep[as can be seen in Fig.~7 in][]{2023A&A...671A..28M}, which would result in the observed thermal pressure \citep{2023A&A...670A..23U}. 
The coexistence of the energy injection of the active galactic nucleus (AGN) into the ICM and the cooling of the ICM in the vicinity of the CGM remains to be understood \citep{2007ARA&A..45..117M,2020NatAs...4...10G,2020IAUS..342...77C, staffehl2025abundanceorigincoolgas, 2025Natur.638..365X}, and NOEMA appears to be a game changer.

\section{Summary and conclusions}
\label{sec:conclusion}

We have presented the first mapping of the tSZ effect in a galaxy cluster with NOEMA. To obtain a clean signal towards \clj, we first removed the point sources in the field, which allowed us to compute the continuum map that represents the unbiased tSZ effect.

We modelled the NOEMA visibilities with a spherical pressure model and, as a result, we have obtained an estimate of the thermal pressure distribution in the cluster core. NOEMA gives a first direct measurement of the pressure in the innermost regions of the ICM, and it favours a flattening of the profile below $\sim 100$~kpc. The pressure profile measured with NOEMA is compatible with previous results within the common radial ranges.

We exploited the complementarity between NOEMA and single-dish instruments \citep{2020EPJWC.22800014L} to obtain a resolved profile that describes the distribution of the thermal pressure in the ICM of \clj\ from $\sim$~15 to $\sim$~1500~kpc. The new measurements
indicate that the cluster has not yet reached a relaxation state.

Our results are highly competitive in comparison with previous works based on ALMA (+ ACA) observations \citep{kitayama2016,Di_Mascolo_2023}. Complementary to ALMA in the Northern Hemisphere, the high angular resolution of NOEMA will be key to resolving high-redshift compact clusters and understanding the co-evolution of gas and galaxies in the very high redshift assembling haloes (clusters or proto-clusters beyond reach of X-ray spectral analyses) that will be detected in the following years in the optical and infrared \citep{2012arXiv1211.0310L, 2016MNRAS.459.1764S}.

\begin{acknowledgements}
This work is based on observations carried out under project number S22BU with the IRAM NOEMA Interferometer. IRAM is supported by INSU/CNRS (France), MPG (Germany) and IGN (Spain). The authors acknowledge IRAM staff for help provided during the observations and for data reduction. We thank C. Romero for providing us with the NIKA, MUSTANG, and Bolocam pressure profile covariance matrices. M.M.E. acknowledges the support of the French Agence Nationale de la Recherche (ANR), under grant ANR-22-CE31-0010. R.A. was supported by the French government through the France 2030 investment plan managed by the National Research Agency (ANR), as part of the Initiative of Excellence of Université Côte d'Azur under reference number ANR-15-IDEX-01. This project was carried out using the Python libraries \texttt{matplotlib} \citep{Hunter2007}, \texttt{numpy} \citep{Harris2020}, \texttt{astropy} \citep{Astropy2013, Astropy2018}, \texttt{chainconsumer} \citep{Hinton2016}, \texttt{emcee}, \texttt{galario}, and \texttt{minot}. We also made use of the IRAM \texttt{GILDAS} software. 
\end{acknowledgements}
\bibliographystyle{aa} 
\bibliography{references_modifs} 

\begin{thebibliography}{75}
\expandafter\ifx\csname natexlab\endcsname\relax\def\natexlab#1{#1}\fi

\bibitem[{{Adam} {et~al.}(2018{\natexlab{a}}){Adam}, {Adane}, {Ade},
  {Andr{\'e}}, {Andrianasolo}, {Aussel}, {Beelen}, {Beno{\^\i}t}, {Bideaud},
  {Billot}, {Bourrion}, {Bracco}, {Calvo}, {Catalano}, {Coiffard}, {Comis}, {De
  Petris}, {D{\'e}sert}, {Doyle}, {Driessen}, {Evans}, {Goupy}, {Kramer},
  {Lagache}, {Leclercq}, {Leggeri}, {Lestrade}, {Mac{\'\i}as-P{\'e}rez},
  {Mauskopf}, {Mayet}, {Maury}, {Monfardini}, {Navarro}, {Pascale}, {Perotto},
  {Pisano}, {Ponthieu}, {Rev{\'e}ret}, {Rigby}, {Ritacco}, {Romero}, {Roussel},
  {Ruppin}, {Schuster}, {Sievers}, {Triqueneaux}, {Tucker}, \&
  {Zylka}}]{adam_nika2_2018}
{Adam}, R., {Adane}, A., {Ade}, P.~A.~R., {et~al.} 2018{\natexlab{a}}, A\&A,
  609, A115

\bibitem[{{Adam} {et~al.}(2015){Adam}, {Comis}, {Mac{\'\i}as-P{\'e}rez},
  {Adane}, {Ade}, {Andr{\'e}}, {Beelen}, {Belier}, {Beno{\^\i}t}, {Bideaud},
  {Billot}, {Blanquer}, {Bourrion}, {Calvo}, {Catalano}, {Coiffard},
  {Cruciani}, {D'Addabbo}, {D{\'e}sert}, {Doyle}, {Goupy}, {Kramer},
  {Leclercq}, {Martino}, {Mauskopf}, {Mayet}, {Monfardini}, {Pajot}, {Pascale},
  {Perotto}, {Pointecouteau}, {Ponthieu}, {Rev{\'e}ret}, {Ritacco},
  {Rodriguez}, {Savini}, {Schuster}, {Sievers}, {Tucker}, \&
  {Zylka}}]{2015A&A...576A..12A}
{Adam}, R., {Comis}, B., {Mac{\'\i}as-P{\'e}rez}, J.~F., {et~al.} 2015, \aap,
  576, A12

\bibitem[{{Adam} {et~al.}(2020){Adam}, {Goksu}, {Leing{\"a}rtner-Goth},
  {Ettori}, {Gnatyk}, {Hnatyk}, {H{\"u}tten}, {P{\'e}rez-Romero},
  {S{\'a}nchez-Conde}, \& {Sergijenko}}]{2020A&A...644A..70A}
{Adam}, R., {Goksu}, H., {Leing{\"a}rtner-Goth}, A., {et~al.} 2020, \aap, 644,
  A70

\bibitem[{{Adam} {et~al.}(2018{\natexlab{b}}){Adam}, {Hahn}, {Ruppin}, {Ade},
  {Andr{\'e}}, {Arnaud}, {Bartalucci}, {Beelen}, {Beno{\^\i}t}, {Bideaud},
  {Billot}, {Bourrion}, {Calvo}, {Catalano}, {Coiffard}, {Comis}, {D'Addabbo},
  {D{\'e}sert}, {Doyle}, {Ferrari}, {Goupy}, {Kramer}, {Lagache}, {Leclercq},
  {Lestrade}, {Mac{\'\i}as-P{\'e}rez}, {Martinez Aviles}, {Martizzi},
  {Maurogordato}, {Mauskopf}, {Mayet}, {Monfardini}, {Pajot}, {Pascale},
  {Perotto}, {Pisano}, {Pointecouteau}, {Ponthieu}, {Pratt}, {Rev{\'e}ret},
  {Ricci}, {Ritacco}, {Rodriguez}, {Romero}, {Roussel}, {Schuster}, {Sievers},
  {Triqueneaux}, {Tucker}, {Wu}, \& {Zylka}}]{2018A&A...614A.118A}
{Adam}, R., {Hahn}, O., {Ruppin}, F., {et~al.} 2018{\natexlab{b}}, \aap, 614,
  A118

\bibitem[{{Adam} {et~al.}(2024){Adam}, {Ricci}, {Eckert}, {Ade}, {Ajeddig},
  {Altieri}, {Andr{\'e}}, {Artis}, {Aussel}, {Beelen}, {Benoist},
  {Beno{\^\i}t}, {Berta}, {Bing}, {Birkinshaw}, {Bourrion}, {Boutigny},
  {Bremer}, {Calvo}, {Cappi}, {Catalano}, {De Petris}, {D{\'e}sert}, {Doyle},
  {Driessen}, {Faccioli}, {Ferrari}, {Gastaldello}, {Giles}, {Gomez}, {Goupy},
  {Hahn}, {Hanser}, {Horellou}, {K{\'e}ruzor{\'e}}, {Koulouridis}, {Kramer},
  {Ladjelate}, {Lagache}, {Leclercq}, {Lestrade}, {Mac{\'\i}as-P{\'e}rez},
  {Madden}, {Maughan}, {Maurogordato}, {Maury}, {Mauskopf}, {Monfardini},
  {Mu{\~n}oz-Echeverr{\'\i}a}, {Pacaud}, {Perotto}, {Pierre}, {Pisano},
  {Pompei}, {Ponthieu}, {Rev{\'e}ret}, {Rigby}, {Ritacco}, {Romero}, {Roussel},
  {Ruppin}, {Sereno}, {Schuster}, {Sievers}, {Tintor{\'e} Vidal}, {Tucker}, \&
  {Zylka}}]{2024A&A...684A..18A}
{Adam}, R., {Ricci}, M., {Eckert}, D., {et~al.} 2024, \aap, 684, A18

\bibitem[{{Allen} {et~al.}(2011){Allen}, {Evrard}, \&
  {Mantz}}]{2011ARA&A..49..409A}
{Allen}, S.~W., {Evrard}, A.~E., \& {Mantz}, A.~B. 2011, \araa, 49, 409

\bibitem[{{Aric{\`o}} {et~al.}(2023){Aric{\`o}}, {Angulo}, {Zennaro},
  {Contreras}, {Chen}, \& {Hern{\'a}ndez-Monteagudo}}]{2023A&A...678A.109A}
{Aric{\`o}}, G., {Angulo}, R.~E., {Zennaro}, M., {et~al.} 2023, \aap, 678, A109

\bibitem[{{Arnaud} {et~al.}(2010){Arnaud}, {Pratt}, {Piffaretti},
  {B{\"o}hringer}, {Croston}, \& {Pointecouteau}}]{2010A&A...517A..92A}
{Arnaud}, M., {Pratt}, G.~W., {Piffaretti}, R., {et~al.} 2010, \aap, 517, A92

\bibitem[{{Astropy Collaboration} {et~al.}(2018){Astropy Collaboration},
  {Price-Whelan}, {Sip{\H{o}}cz}, {G{\"u}nther}, {Lim}, {Crawford}, {Conseil},
  {Shupe}, {Craig}, {Dencheva}, {Ginsburg}, {VanderPlas}, {Bradley},
  {P{\'e}rez-Su{\'a}rez}, {de Val-Borro}, {Aldcroft}, {Cruz}, {Robitaille},
  {Tollerud}, {Ardelean}, {Babej}, {Bach}, {Bachetti}, {Bakanov}, {Bamford},
  {Barentsen}, {Barmby}, {Baumbach}, {Berry}, {Biscani}, {Boquien}, {Bostroem},
  {Bouma}, {Brammer}, {Bray}, {Breytenbach}, {Buddelmeijer}, {Burke},
  {Calderone}, {Cano Rodr{\'\i}guez}, {Cara}, {Cardoso}, {Cheedella}, {Copin},
  {Corrales}, {Crichton}, {D'Avella}, {Deil}, {Depagne}, {Dietrich}, {Donath},
  {Droettboom}, {Earl}, {Erben}, {Fabbro}, {Ferreira}, {Finethy}, {Fox},
  {Garrison}, {Gibbons}, {Goldstein}, {Gommers}, {Greco}, {Greenfield},
  {Groener}, {Grollier}, {Hagen}, {Hirst}, {Homeier}, {Horton}, {Hosseinzadeh},
  {Hu}, {Hunkeler}, {Ivezi{\'c}}, {Jain}, {Jenness}, {Kanarek}, {Kendrew},
  {Kern}, {Kerzendorf}, {Khvalko}, {King}, {Kirkby}, {Kulkarni}, {Kumar},
  {Lee}, {Lenz}, {Littlefair}, {Ma}, {Macleod}, {Mastropietro}, {McCully},
  {Montagnac}, {Morris}, {Mueller}, {Mumford}, {Muna}, {Murphy}, {Nelson},
  {Nguyen}, {Ninan}, {N{\"o}the}, {Ogaz}, {Oh}, {Parejko}, {Parley}, {Pascual},
  {Patil}, {Patil}, {Plunkett}, {Prochaska}, {Rastogi}, {Reddy Janga},
  {Sabater}, {Sakurikar}, {Seifert}, {Sherbert}, {Sherwood-Taylor}, {Shih},
  {Sick}, {Silbiger}, {Singanamalla}, {Singer}, {Sladen}, {Sooley},
  {Sornarajah}, {Streicher}, {Teuben}, {Thomas}, {Tremblay}, {Turner},
  {Terr{\'o}n}, {van Kerkwijk}, {de la Vega}, {Watkins}, {Weaver}, {Whitmore},
  {Woillez}, {Zabalza}, \& {Astropy Contributors}}]{Astropy2018}
{Astropy Collaboration}, {Price-Whelan}, A.~M., {Sip{\H{o}}cz}, B.~M., {et~al.}
  2018, \aj, 156, 123

\bibitem[{{Astropy Collaboration} {et~al.}(2013){Astropy Collaboration},
  {Robitaille}, {Tollerud}, {Greenfield}, {Droettboom}, {Bray}, {Aldcroft},
  {Davis}, {Ginsburg}, {Price-Whelan}, {Kerzendorf}, {Conley}, {Crighton},
  {Barbary}, {Muna}, {Ferguson}, {Grollier}, {Parikh}, {Nair}, {Unther},
  {Deil}, {Woillez}, {Conseil}, {Kramer}, {Turner}, {Singer}, {Fox}, {Weaver},
  {Zabalza}, {Edwards}, {Azalee Bostroem}, {Burke}, {Casey}, {Crawford},
  {Dencheva}, {Ely}, {Jenness}, {Labrie}, {Lim}, {Pierfederici}, {Pontzen},
  {Ptak}, {Refsdal}, {Servillat}, \& {Streicher}}]{Astropy2013}
{Astropy Collaboration}, {Robitaille}, T.~P., {Tollerud}, E.~J., {et~al.} 2013,
  \aap, 558, A33

\bibitem[{{Basu} {et~al.}(2016){Basu}, {Sommer}, {Erler}, {Eckert}, {Vazza},
  {Magnelli}, {Bertoldi}, \& {Tozzi}}]{basu2016}
{Basu}, K., {Sommer}, M., {Erler}, J., {et~al.} 2016, ApJL, 829, L23

\bibitem[{{Basu} {et~al.}(2010){Basu}, {Zhang}, {Sommer}, {Bender}, {Bertoldi},
  {Dobbs}, {Eckmiller}, {Halverson}, {Holzapfel}, {Horellou}, {Jaritz},
  {Johansson}, {Johnson}, {Kennedy}, {Kneissl}, {Lanting}, {Lee}, {Mehl},
  {Menten}, {Navarrete}, {Pacaud}, {Reichardt}, {Reiprich}, {Richards},
  {Schwan}, \& {Westbrook}}]{2010A&A...519A..29B}
{Basu}, K., {Zhang}, Y.~Y., {Sommer}, M.~W., {et~al.} 2010, \aap, 519, A29

\bibitem[{{Bourrion} {et~al.}(2016){Bourrion}, {Benoit}, {Bouly}, {Bouvier},
  {Bosson}, {Calvo}, {Catalano}, {Goupy}, {Li}, {Mac{\'\i}as-P{\'e}rez},
  {Monfardini}, {Tourres}, {Ponchant}, \& {Vescovi}}]{NIKA2-electronics}
{Bourrion}, O., {Benoit}, A., {Bouly}, J.~L., {et~al.} 2016, Journal of
  Instrumentation, 11, P11001

\bibitem[{{Calvo} {et~al.}(2016){Calvo}, {Beno{\^\i}t}, {Catalano}, {Goupy},
  {Monfardini}, {Ponthieu}, {Barria}, {Bres}, {Grollier}, {Garde}, {Leggeri},
  {Pont}, {Triqueneaux}, {Adam}, {Bourrion}, {Mac{\'\i}as-P{\'e}rez}, {Rebolo},
  {Ritacco}, {Scordilis}, {Tourres}, {Adane}, {Coiffard}, {Leclercq},
  {D{\'e}sert}, {Doyle}, {Mauskopf}, {Tucker}, {Ade}, {Andr{\'e}}, {Beelen},
  {Belier}, {Bideaud}, {Billot}, {Comis}, {D'Addabbo}, {Kramer}, {Martino},
  {Mayet}, {Pajot}, {Pascale}, {Perotto}, {Rev{\'e}ret}, {Ritacco},
  {Rodriguez}, {Savini}, {Schuster}, {Sievers}, \& {Zylka}}]{calvo}
{Calvo}, M., {Beno{\^\i}t}, A., {Catalano}, A., {et~al.} 2016, Journal of Low
  Temperature Physics, 184, 816

\bibitem[{{Carlstrom} {et~al.}(2002){Carlstrom}, {Holder}, \&
  {Reese}}]{carlstrom2002}
{Carlstrom}, J.~E., {Holder}, G.~P., \& {Reese}, E.~D. 2002, ARA\&A, 40, 643

\bibitem[{{Carlstrom} {et~al.}(1996){Carlstrom}, {Joy}, \&
  {Grego}}]{carlstrom1996}
{Carlstrom}, J.~E., {Joy}, M., \& {Grego}, L. 1996, ApJL, 456, L75

\bibitem[{{Clark}(1980)}]{1980A&A....89..377C}
{Clark}, B.~G. 1980, \aap, 89, 377

\bibitem[{{Combes}(2020)}]{2020IAUS..342...77C}
{Combes}, F. 2020, in IAU Symposium, Vol. 342, Perseus in Sicily: From Black
  Hole to Cluster Outskirts, ed. K.~{Asada}, E.~{de Gouveia Dal Pino},
  M.~{Giroletti}, H.~{Nagai}, \& R.~{Nemmen}, 77--84

\bibitem[{{Cusworth} {et~al.}(2014){Cusworth}, {Kay}, {Battye}, \&
  {Thomas}}]{2014MNRAS.439.2485C}
{Cusworth}, S.~J., {Kay}, S.~T., {Battye}, R.~A., \& {Thomas}, P.~A. 2014,
  \mnras, 439, 2485

\bibitem[{{Di Gennaro} {et~al.}(2021){Di Gennaro}, {van Weeren}, {Brunetti},
  {Cassano}, {Br{\"u}ggen}, {Hoeft}, {Shimwell}, {R{\"o}ttgering}, {Bonafede},
  {Botteon}, {Cuciti}, {Dallacasa}, {de Gasperin},
  {Dom{\'\i}nguez-Fern{\'a}ndez}, {En{\ss}lin}, {Gastaldello}, {Mandal},
  {Rossetti}, \& {Simionescu}}]{2021NatAs...5..268D}
{Di Gennaro}, G., {van Weeren}, R.~J., {Brunetti}, G., {et~al.} 2021, Nature
  Astronomy, 5, 268

\bibitem[{{Di Mascolo} {et~al.}(2023){Di Mascolo}, {Saro}, {Mroczkowski},
  {Borgani}, {Churazov}, {Rasia}, {Tozzi}, {Dannerbauer}, {Basu}, {Carilli},
  {Ginolfi}, {Miley}, {Nonino}, {Pannella}, {Pentericci}, \&
  {Rizzo}}]{Di_Mascolo_2023}
{Di Mascolo}, L., {Saro}, A., {Mroczkowski}, T., {et~al.} 2023, Nature, 615,
  809

\bibitem[{{Dicker} {et~al.}(2014){Dicker}, {Ade}, {Aguirre}, {Brevik}, {Cho},
  {Datta}, {Devlin}, {Dober}, {Egan}, {Ford}, {Ford}, {Hilton}, {Hubmayr},
  {Irwin}, {Mason}, {Marganian}, {Mello}, {McMahon}, {Mroczkowski}, {Romero},
  {Stanchfield}, {Tucker}, {Vale}, {White}, {Whitehead}, \&
  {Young}}]{dicker_mustang2_2014}
{Dicker}, S.~R., {Ade}, P.~A.~R., {Aguirre}, J., {et~al.} 2014, in Society of
  Photo-Optical Instrumentation Engineers (SPIE) Conference Series, Vol. 9153,
  Millimeter, Submillimeter, and Far-Infrared Detectors and Instrumentation for
  Astronomy VII, ed. W.~S. {Holland} \& J.~{Zmuidzinas}, 91530J

\bibitem[{{Ebeling} {et~al.}(2001){Ebeling}, {Jones}, {Fairley}, {Perlman},
  {Scharf}, \& {Horner}}]{2001ApJ...548L..23E}
{Ebeling}, H., {Jones}, L.~R., {Fairley}, B.~W., {et~al.} 2001, \apjl, 548, L23

\bibitem[{{Foreman-Mackey} {et~al.}(2019){Foreman-Mackey}, {Farr}, {Sinha},
  {Archibald}, {Hogg}, {Sanders}, {Zuntz}, {Williams}, {Nelson}, {de
  Val-Borro}, {Erhardt}, {Pashchenko}, \& {Pla}}]{2019JOSS....4.1864F}
{Foreman-Mackey}, D., {Farr}, W., {Sinha}, M., {et~al.} 2019, The Journal of
  Open Source Software, 4, 1864

\bibitem[{{Gaspari} {et~al.}(2020){Gaspari}, {Tombesi}, \&
  {Cappi}}]{2020NatAs...4...10G}
{Gaspari}, M., {Tombesi}, F., \& {Cappi}, M. 2020, Nature Astronomy, 4, 10

\bibitem[{{Gelman} \& {Rubin}(1992)}]{gelmanrubin}
{Gelman}, A. \& {Rubin}, D.~B. 1992, Statistical Science, 7, 457

\bibitem[{{Goodman} \& {Weare}(2010)}]{2010CAMCS...5...65G}
{Goodman}, J. \& {Weare}, J. 2010, Communications in Applied Mathematics and
  Computational Science, 5, 65

\bibitem[{{Grainge} {et~al.}(1993){Grainge}, {Jones}, {Pooley}, {Saunders}, \&
  {Edge}}]{grainge1993}
{Grainge}, K., {Jones}, M., {Pooley}, G., {Saunders}, R., \& {Edge}, A. 1993,
  MNRAS, 265, L57

\bibitem[{Harris {et~al.}(2020)Harris, Millman, van~der Walt, Gommers,
  Virtanen, Cournapeau, Wieser, Taylor, Berg, Smith, Kern, Picus, Hoyer, van
  Kerkwijk, Brett, Haldane, del R{\'{i}}o, Wiebe, Peterson,
  G{\'{e}}rard-Marchant, Sheppard, Reddy, Weckesser, Abbasi, Gohlke, \&
  Oliphant}]{Harris2020}
Harris, C.~R., Millman, K.~J., van~der Walt, S.~J., {et~al.} 2020, Nature, 585,
  357

\bibitem[{{Hinton}(2016)}]{Hinton2016}
{Hinton}, S.~R. 2016, The Journal of Open Source Software, 1, 00045

\bibitem[{{Holden} {et~al.}(2009){Holden}, {Franx}, {Illingworth}, {Postman},
  {van der Wel}, {Kelson}, {Blakeslee}, {Ford}, {Demarco}, \&
  {Mei}}]{2009ApJ...693..617H}
{Holden}, B.~P., {Franx}, M., {Illingworth}, G.~D., {et~al.} 2009, \apj, 693,
  617

\bibitem[{Hunter(2007)}]{Hunter2007}
Hunter, J.~D. 2007, Computing in Science \& Engineering, 9, 90

\bibitem[{{Jee} \& {Tyson}(2009)}]{2009ApJ...691.1337J}
{Jee}, M.~J. \& {Tyson}, J.~A. 2009, \apj, 691, 1337

\bibitem[{{Jones} {et~al.}(1993){Jones}, {Saunders}, {Alexander}, {Birkinshaw},
  {Dillon}, {Grainge}, {Hancock}, {Lasenby}, {Lefebvre}, {Pooley}, {Scott},
  {Titterington}, \& {Wilson}}]{jones1993}
{Jones}, M., {Saunders}, R., {Alexander}, P., {et~al.} 1993, Nature, 365, 320

\bibitem[{{Joy} {et~al.}(2001){Joy}, {LaRoque}, {Grego}, {Carlstrom}, {Dawson},
  {Ebeling}, {Holzapfel}, {Nagai}, \& {Reese}}]{2001ApJ...551L...1J}
{Joy}, M., {LaRoque}, S., {Grego}, L., {et~al.} 2001, \apjl, 551, L1

\bibitem[{{K{\'e}ruzor{\'e}} {et~al.}(2023){K{\'e}ruzor{\'e}}, {Mayet},
  {Artis}, {Mac{\'\i}as-P{\'e}rez}, {Mu{\~n}oz-Echeverr{\'\i}a}, {Perotto}, \&
  {Ruppin}}]{2023OJAp....6E...9K}
{K{\'e}ruzor{\'e}}, F., {Mayet}, F., {Artis}, E., {et~al.} 2023, The Open
  Journal of Astrophysics, 6, 9

\bibitem[{{Kitayama} {et~al.}(2016){Kitayama}, {Ueda}, {Takakuwa}, {Tsutsumi},
  {Komatsu}, {Akahori}, {Iono}, {Izumi}, {Kawabe}, {Kohno}, {Matsuo}, {Ota},
  {Suto}, {Takizawa}, \& {Yoshikawa}}]{kitayama2016}
{Kitayama}, T., {Ueda}, S., {Takakuwa}, S., {et~al.} 2016, PASJ, 68, 88

\bibitem[{{Korngut} {et~al.}(2011){Korngut}, {Dicker}, {Reese}, {Mason},
  {Devlin}, {Mroczkowski}, {Sarazin}, {Sun}, \&
  {Sievers}}]{2011ApJ...734...10K}
{Korngut}, P.~M., {Dicker}, S.~R., {Reese}, E.~D., {et~al.} 2011, \apj, 734, 10

\bibitem[{{Lef{\`e}vre} {et~al.}(2020){Lef{\`e}vre}, {Kramer}, {Neri}, {Berta},
  \& {Schuster}}]{2020EPJWC.22800014L}
{Lef{\`e}vre}, C., {Kramer}, C., {Neri}, R., {Berta}, S., \& {Schuster}, K.
  2020, in European Physical Journal Web of Conferences, Vol. 228, mm Universe
  @ NIKA2 - Observing the mm Universe with the NIKA2 Camera, 00014

\bibitem[{{Lepore} {et~al.}(2024){Lepore}, {Di Mascolo}, {Tozzi}, {Churazov},
  {Mroczkowski}, {Borgani}, {Carilli}, {Gaspari}, {Ginolfi}, {Liu},
  {Pentericci}, {Rasia}, {Rosati}, {R{\"o}ttgering}, {Anderson}, {Dannerbauer},
  {Miley}, \& {Norman}}]{2024A&A...682A...186L}
{Lepore}, M., {Di Mascolo}, L., {Tozzi}, P., {et~al.} 2024, \aap, 682, A186

\bibitem[{{LSST Dark Energy Science Collaboration}(2012)}]{2012arXiv1211.0310L}
{LSST Dark Energy Science Collaboration}. 2012, arXiv e-prints, arXiv:1211.0310

\bibitem[{{Maughan} {et~al.}(2007){Maughan}, {Jones}, {Jones}, \& {Van
  Speybroeck}}]{2007ApJ...659.1125M}
{Maughan}, B.~J., {Jones}, C., {Jones}, L.~R., \& {Van Speybroeck}, L. 2007,
  \apj, 659, 1125

\bibitem[{{Mayet} {et~al.}(2020){Mayet}, {Adam}, {Ade}, {Andr{\'e}},
  {Andrianasolo}, {Arnaud}, {Aussel}, {Bartalucci}, {Beelen}, {Beno{\^\i}t},
  {Bideaud}, {Bourrion}, {Calvo}, {Catalano}, {Comis}, {De Petris},
  {D{\'e}sert}, {Doyle}, {Driessen}, {Gomez}, {Goupy}, {K{\'e}ruzor{\'e}},
  {Kramer}, {Ladjelate}, {Lagache}, {Leclercq}, {Lestrade},
  {Mac{\'\i}as-P{\'e}rez}, {Mauskopf}, {Monfardini}, {Perotto}, {Pisano},
  {Pointecouteau}, {Ponthieu}, {Pratt}, {Rev{\'e}ret}, {Ritacco}, {Romero},
  {Roussel}, {Ruppin}, {Schuster}, {Shu}, {Sievers}, {Tucker}, \&
  {Zylka}}]{2020EPJWC.22800017M}
{Mayet}, F., {Adam}, R., {Ade}, P., {et~al.} 2020, in European Physical Journal
  Web of Conferences, Vol. 228, mm Universe @ NIKA2 - Observing the mm Universe
  with the NIKA2 Camera, 00017

\bibitem[{{McNamara} \& {Nulsen}(2007)}]{2007ARA&A..45..117M}
{McNamara}, B.~R. \& {Nulsen}, P.~E.~J. 2007, \araa, 45, 117

\bibitem[{{Melin} \& {Pratt}(2023)}]{2023A&A...678A.197M}
{Melin}, J.~B. \& {Pratt}, G.~W. 2023, \aap, 678, A197

\bibitem[{{Moffet} \& {Birkinshaw}(1989)}]{Moffet&Birkinshaw1989}
{Moffet}, A.~T. \& {Birkinshaw}, M. 1989, AJ, 98, 1148

\bibitem[{{Mu{\~n}oz-Echeverr{\'\i}a}
  {et~al.}(2023){Mu{\~n}oz-Echeverr{\'\i}a}, {Mac{\'\i}as-P{\'e}rez}, {Pratt},
  {Adam}, {Ade}, {Ajeddig}, {Andr{\'e}}, {Arnaud}, {Artis}, {Aussel},
  {Bartalucci}, {Beelen}, {Beno{\^\i}t}, {Berta}, {Bing}, {Bourrion}, {Calvo},
  {Catalano}, {De Petris}, {D{\'e}sert}, {Doyle}, {Driessen}, {Ferragamo},
  {Gomez}, {Goupy}, {Hanser}, {K{\'e}ruzor{\'e}}, {Kramer}, {Ladjelate},
  {Lagache}, {Leclercq}, {Lestrade}, {Maury}, {Mauskopf}, {Mayet}, {Melin},
  {Monfardini}, {Paliwal}, {Perotto}, {Pisano}, {Pointecouteau}, {Ponthieu},
  {Rev{\'e}ret}, {Rigby}, {Ritacco}, {Romero}, {Roussel}, {Ruppin}, {Schuster},
  {Shu}, {Sievers}, {Tucker}, \& {Yepes}}]{2023A&A...671A..28M}
{Mu{\~n}oz-Echeverr{\'\i}a}, M., {Mac{\'\i}as-P{\'e}rez}, J.~F., {Pratt},
  G.~W., {et~al.} 2023, \aap, 671, A28

\bibitem[{{Muchovej} {et~al.}(2007){Muchovej}, {Mroczkowski}, {Carlstrom},
  {Cartwright}, {Greer}, {Hennessy}, {Loh}, {Pryke}, {Reddall}, {Runyan},
  {Sharp}, {Hawkins}, {Lamb}, {Woody}, {Joy}, {Leitch}, \& {Miller}}]{muchovej}
{Muchovej}, S., {Mroczkowski}, T., {Carlstrom}, J.~E., {et~al.} 2007, ApJ, 663,
  708

\bibitem[{{Nagai} {et~al.}(2007){Nagai}, {Kravtsov}, \&
  {Vikhlinin}}]{2007ApJ...668....1N}
{Nagai}, D., {Kravtsov}, A.~V., \& {Vikhlinin}, A. 2007, \apj, 668, 1

\bibitem[{{Neri} \& {S{\'a}nchez-Portal}(2023)}]{2023pcsf.conf..308N}
{Neri}, R. \& {S{\'a}nchez-Portal}, M. 2023, in Physics and Chemistry of Star
  Formation: The Dynamical ISM Across Time and Spatial Scales, ed.
  V.~{Ossenkopf-Okada}, R.~{Schaaf}, I.~{Breloy}, \& J.~{Stutzki}, 308

\bibitem[{{Perotto} {et~al.}(2022){Perotto}, {Adam}, {Ade}, {Ajeddig},
  {Andr{\'e}}, {Arnaud}, {Artis}, {Aussel}, {Bartalucci}, {Beelen},
  {Beno{\^\i}t}, {Berta}, {Bing}, {Bourrion}, {Calvo}, {Catalano}, {De Petris},
  {D{\'e}sert}, {Doyle}, {Driessen}, {Ferragamo}, {Gomez}, {Goupy},
  {K{\'e}ruzor{\'e}}, {Kramer}, {Ladjelate}, {Lagache}, {Leclercq}, {Lestrade},
  {Mac{\'\i}as-P{\'e}rez}, {Maury}, {Mauskopf}, {Mayet}, {Monfardini},
  {Mu{\~n}oz-Echeverr{\'\i}a}, {Paliwal}, {Pisano}, {Pointecouteau},
  {Ponthieu}, {Pratt}, {Rev{\'e}ret}, {Rigby}, {Ritacco}, {Romero}, {Roussel},
  {Ruppin}, {Schuster}, {Shu}, {Sievers}, {Tucker}, \&
  {Yepes}}]{2022EPJWC.25700038P}
{Perotto}, L., {Adam}, R., {Ade}, P., {et~al.} 2022, in European Physical
  Journal Web of Conferences, Vol. 257, mm Universe @ NIKA2 - Observing the mm
  Universe with the NIKA2 Camera, 00038

\bibitem[{{Perotto} {et~al.}(2020){Perotto}, {Ponthieu},
  {Mac{\'\i}as-P{\'e}rez}, {Adam}, {Ade}, {Andr{\'e}}, {Andrianasolo},
  {Aussel}, {Beelen}, {Beno{\^\i}t}, {Berta}, {Bideaud}, {Bourrion}, {Calvo},
  {Catalano}, {Comis}, {De Petris}, {D{\'e}sert}, {Doyle}, {Driessen},
  {Garc{\'\i}a}, {Gomez}, {Goupy}, {John}, {K{\'e}ruzor{\'e}}, {Kramer},
  {Ladjelate}, {Lagache}, {Leclercq}, {Lestrade}, {Maury}, {Mauskopf}, {Mayet},
  {Monfardini}, {Navarro}, {Pe{\~n}alver}, {Pierfederici}, {Pisano},
  {Rev{\'e}ret}, {Ritacco}, {Romero}, {Roussel}, {Ruppin}, {Schuster}, {Shu},
  {Sievers}, {Tucker}, \& {Zylka}}]{2020A&A...637A..71P}
{Perotto}, L., {Ponthieu}, N., {Mac{\'\i}as-P{\'e}rez}, J.~F., {et~al.} 2020,
  \aap, 637, A71

\bibitem[{{Planck Collaboration} {et~al.}(2013){Planck Collaboration}, {Ade},
  {Aghanim}, {Arnaud}, {Ashdown}, {Atrio-Barandela}, {Aumont}, {Baccigalupi},
  {Balbi}, {Banday}, {Barreiro}, {Bartlett}, {Battaner}, {Benabed},
  {Beno{\^\i}t}, {Bernard}, {Bersanelli}, {Bhatia}, {Bikmaev}, {Bobin},
  {B{\"o}hringer}, {Bonaldi}, {Bond}, {Borgani}, {Borrill}, {Bouchet},
  {Bourdin}, {Brown}, {Burenin}, {Burigana}, {Cabella}, {Cardoso}, {Carvalho},
  {Castex}, {Catalano}, {Cay{\'o}n}, {Chamballu}, {Chiang}, {Chon},
  {Christensen}, {Churazov}, {Clements}, {Colafrancesco}, {Colombi}, {Colombo},
  {Comis}, {Coulais}, {Crill}, {Cuttaia}, {Da Silva}, {Dahle}, {Danese},
  {Davis}, {de Bernardis}, {de Gasperis}, {de Zotti}, {Delabrouille},
  {D{\'e}mocl{\`e}s}, {D{\'e}sert}, {Diego}, {Dolag}, {Dole}, {Donzelli},
  {Dor{\'e}}, {D{\"o}rl}, {Douspis}, {Dupac}, {Efstathiou}, {En{\ss}lin},
  {Eriksen}, {Finelli}, {Flores-Cacho}, {Forni}, {Fosalba}, {Frailis},
  {Franceschi}, {Frommert}, {Galeotta}, {Ganga}, {G{\'e}nova-Santos}, {Giard},
  {Giraud-H{\'e}raud}, {Gonz{\'a}lez-Nuevo}, {G{\'o}rski}, {Gregorio},
  {Gruppuso}, {Hansen}, {Harrison}, {Hempel}, {Henrot-Versill{\'e}},
  {Hern{\'a}ndez-Monteagudo}, {Herranz}, {Hildebrandt}, {Hivon}, {Hobson},
  {Holmes}, {Hurier}, {Jaffe}, {Jaffe}, {Jagemann}, {Jones}, {Juvela},
  {Keih{\"a}nen}, {Khamitov}, {Kisner}, {Kneissl}, {Knoche}, {Knox}, {Kunz},
  {Kurki-Suonio}, {Lagache}, {L{\"a}hteenm{\"a}ki}, {Lamarre}, {Lasenby},
  {Lawrence}, {Le Jeune}, {Leonardi}, {Liddle}, {Lilje}, {L{\'o}pez-Caniego},
  {Luzzi}, {Mac{\'\i}as-P{\'e}rez}, {Maino}, {Mandolesi}, {Maris}, {Marleau},
  {Marshall}, {Mart{\'\i}nez-Gonz{\'a}lez}, {Masi}, {Massardi}, {Matarrese},
  {Mazzotta}, {Mei}, {Melchiorri}, {Melin}, {Mendes}, {Mennella}, {Mitra},
  {Miville-Desch{\^e}nes}, {Moneti}, {Montier}, {Morgante}, {Mortlock},
  {Munshi}, {Murphy}, {Naselsky}, {Nati}, {Natoli}, {N{\o}rgaard-Nielsen},
  {Noviello}, {Novikov}, {Novikov}, {Osborne}, {Pajot}, {Paoletti}, {Pasian},
  {Patanchon}, {Perdereau}, {Perotto}, {Perrotta}, {Piacentini}, {Piat},
  {Pierpaoli}, {Piffaretti}, {Plaszczynski}, {Pointecouteau}, {Polenta},
  {Ponthieu}, {Popa}, {Poutanen}, {Pratt}, {Prunet}, {Puget}, {Rachen},
  {Reach}, {Rebolo}, {Reinecke}, {Remazeilles}, {Renault}, {Ricciardi},
  {Riller}, {Ristorcelli}, {Rocha}, {Roman}, {Rosset}, {Rossetti},
  {Rubi{\~n}o-Mart{\'\i}n}, {Rusholme}, {Sandri}, {Savini}, {Scott}, {Smoot},
  {Starck}, {Sudiwala}, {Sunyaev}, {Sutton}, {Suur-Uski}, {Sygnet}, {Tauber},
  {Terenzi}, {Toffolatti}, {Tomasi}, {Tristram}, {Tuovinen}, {Valenziano}, {Van
  Tent}, {Varis}, {Vielva}, {Villa}, {Vittorio}, {Wade}, {Wandelt}, {Welikala},
  {White}, {White}, {Yvon}, {Zacchei}, \& {Zonca}}]{2013A&A...550A.131P}
{Planck Collaboration}, {Ade}, P.~A.~R., {Aghanim}, N., {et~al.} 2013, \aap,
  550, A131

\bibitem[{{Planck Collaboration} {et~al.}(2016){Planck Collaboration}, {Ade},
  {Aghanim}, {Arnaud}, {Ashdown}, {Aumont}, {Baccigalupi}, {Banday},
  {Barreiro}, {Barrena}, {Bartlett}, {Bartolo}, {Battaner}, {Battye},
  {Benabed}, {Beno{\^\i}t}, {Benoit-L{\'e}vy}, {Bernard}, {Bersanelli},
  {Bielewicz}, {Bikmaev}, {B{\"o}hringer}, {Bonaldi}, {Bonavera}, {Bond},
  {Borrill}, {Bouchet}, {Bucher}, {Burenin}, {Burigana}, {Butler}, {Calabrese},
  {Cardoso}, {Carvalho}, {Catalano}, {Challinor}, {Chamballu}, {Chary},
  {Chiang}, {Chon}, {Christensen}, {Clements}, {Colombi}, {Colombo}, {Combet},
  {Comis}, {Couchot}, {Coulais}, {Crill}, {Curto}, {Cuttaia}, {Dahle},
  {Danese}, {Davies}, {Davis}, {de Bernardis}, {de Rosa}, {de Zotti},
  {Delabrouille}, {D{\'e}sert}, {Dickinson}, {Diego}, {Dolag}, {Dole},
  {Donzelli}, {Dor{\'e}}, {Douspis}, {Ducout}, {Dupac}, {Efstathiou},
  {Eisenhardt}, {Elsner}, {En{\ss}lin}, {Eriksen}, {Falgarone}, {Fergusson},
  {Feroz}, {Ferragamo}, {Finelli}, {Forni}, {Frailis}, {Fraisse}, {Franceschi},
  {Frejsel}, {Galeotta}, {Galli}, {Ganga}, {G{\'e}nova-Santos}, {Giard},
  {Giraud-H{\'e}raud}, {Gjerl{\o}w}, {Gonz{\'a}lez-Nuevo}, {G{\'o}rski},
  {Grainge}, {Gratton}, {Gregorio}, {Gruppuso}, {Gudmundsson}, {Hansen},
  {Hanson}, {Harrison}, {Hempel}, {Henrot-Versill{\'e}},
  {Hern{\'a}ndez-Monteagudo}, {Herranz}, {Hildebrandt}, {Hivon}, {Hobson},
  {Holmes}, {Hornstrup}, {Hovest}, {Huffenberger}, {Hurier}, {Jaffe}, {Jaffe},
  {Jin}, {Jones}, {Juvela}, {Keih{\"a}nen}, {Keskitalo}, {Khamitov}, {Kisner},
  {Kneissl}, {Knoche}, {Kunz}, {Kurki-Suonio}, {Lagache}, {Lamarre}, {Lasenby},
  {Lattanzi}, {Lawrence}, {Leonardi}, {Lesgourgues}, {Levrier}, {Liguori},
  {Lilje}, {Linden-V{\o}rnle}, {L{\'o}pez-Caniego}, {Lubin},
  {Mac{\'\i}as-P{\'e}rez}, {Maggio}, {Maino}, {Mak}, {Mandolesi}, {Mangilli},
  {Martin}, {Mart{\'\i}nez-Gonz{\'a}lez}, {Masi}, {Matarrese}, {Mazzotta},
  {McGehee}, {Mei}, {Melchiorri}, {Melin}, {Mendes}, {Mennella}, {Migliaccio},
  {Mitra}, {Miville-Desch{\^e}nes}, {Moneti}, {Montier}, {Morgante},
  {Mortlock}, {Moss}, {Munshi}, {Murphy}, {Naselsky}, {Nastasi}, {Nati},
  {Natoli}, {Netterfield}, {N{\o}rgaard-Nielsen}, {Noviello}, {Novikov},
  {Novikov}, {Olamaie}, {Oxborrow}, {Paci}, {Pagano}, {Pajot}, {Paoletti},
  {Pasian}, {Patanchon}, {Pearson}, {Perdereau}, {Perotto}, {Perrott},
  {Perrotta}, {Pettorino}, {Piacentini}, {Piat}, {Pierpaoli}, {Pietrobon},
  {Plaszczynski}, {Pointecouteau}, {Polenta}, {Pratt}, {Pr{\'e}zeau}, {Prunet},
  {Puget}, {Rachen}, {Reach}, {Rebolo}, {Reinecke}, {Remazeilles}, {Renault},
  {Renzi}, {Ristorcelli}, {Rocha}, {Rosset}, {Rossetti}, {Roudier}, {Rozo},
  {Rubi{\~n}o-Mart{\'\i}n}, {Rumsey}, {Rusholme}, {Rykoff}, {Sandri}, {Santos},
  {Saunders}, {Savelainen}, {Savini}, {Schammel}, {Scott}, {Seiffert},
  {Shellard}, {Shimwell}, {Spencer}, {Stanford}, {Stern}, {Stolyarov},
  {Stompor}, {Streblyanska}, {Sudiwala}, {Sunyaev}, {Sutton}, {Suur-Uski},
  {Sygnet}, {Tauber}, {Terenzi}, {Toffolatti}, {Tomasi}, {Tramonte},
  {Tristram}, {Tucci}, {Tuovinen}, {Umana}, {Valenziano}, {Valiviita}, {Van
  Tent}, {Vielva}, {Villa}, {Wade}, {Wandelt}, {Wehus}, {White}, {Wright},
  {Yvon}, {Zacchei}, \& {Zonca}}]{2016A&A...594A..27P}
{Planck Collaboration}, {Ade}, P.~A.~R., {Aghanim}, N., {et~al.} 2016, \aap,
  594, A27

\bibitem[{{Planck Collaboration XVIII.}(2015)}]{planck2016}
{Planck Collaboration XVIII.} 2015, A\&A, 594, A27

\bibitem[{{Romero} {et~al.}(2018){Romero}, {McWilliam},
  {Mac{\'\i}as-P{\'e}rez}, {Adam}, {Ade}, {Andr{\'e}}, {Aussel}, {Beelen},
  {Beno{\^\i}t}, {Bideaud}, {Billot}, {Bourrion}, {Calvo}, {Catalano},
  {Coiffard}, {Comis}, {de Petris}, {D{\'e}sert}, {Doyle}, {Goupy}, {Kramer},
  {Lagache}, {Leclercq}, {Lestrade}, {Mauskopf}, {Mayet}, {Monfardini},
  {Pascale}, {Perotto}, {Pisano}, {Ponthieu}, {Rev{\'e}ret}, {Ritacco},
  {Roussel}, {Ruppin}, {Schuster}, {Sievers}, {Triqueneaux}, {Tucker}, \&
  {Zylka}}]{2018A&A...612A..39R}
{Romero}, C., {McWilliam}, M., {Mac{\'\i}as-P{\'e}rez}, J.~F., {et~al.} 2018,
  \aap, 612, A39

\bibitem[{{Romero} {et~al.}(2017){Romero}, {Mason}, {Sayers}, {Mroczkowski},
  {Sarazin}, {Donahue}, {Baldi}, {Clarke}, {Young}, {Sievers}, {Dicker},
  {Reese}, {Czakon}, {Devlin}, {Korngut}, \& {Golwala}}]{2017ApJ...838...86R}
{Romero}, C.~E., {Mason}, B.~S., {Sayers}, J., {et~al.} 2017, \apj, 838, 86

\bibitem[{{Ruppin} {et~al.}(2017){Ruppin}, {Adam}, {Comis}, {Ade}, {Andr{\'e}},
  {Arnaud}, {Beelen}, {Beno{\^\i}t}, {Bideaud}, {Billot}, {Bourrion}, {Calvo},
  {Catalano}, {Coiffard}, {D'Addabbo}, {De Petris}, {D{\'e}sert}, {Doyle},
  {Goupy}, {Kramer}, {Leclercq}, {Mac{\'\i}as-P{\'e}rez}, {Mauskopf}, {Mayet},
  {Monfardini}, {Pajot}, {Pascale}, {Perotto}, {Pisano}, {Pointecouteau},
  {Ponthieu}, {Pratt}, {Rev{\'e}ret}, {Ritacco}, {Rodriguez}, {Romero},
  {Schuster}, {Sievers}, {Triqueneaux}, {Tucker}, \&
  {Zylka}}]{2017A&A...597A.110R}
{Ruppin}, F., {Adam}, R., {Comis}, B., {et~al.} 2017, \aap, 597, A110

\bibitem[{{Ruppin} {et~al.}(2018){Ruppin}, {Mayet}, {Pratt}, {Adam}, {Ade},
  {Andr{\'e}}, {Arnaud}, {Aussel}, {Bartalucci}, {Beelen}, {Beno{\^\i}t},
  {Bideaud}, {Bourrion}, {Calvo}, {Catalano}, {Comis}, {De Petris},
  {D{\'e}sert}, {Doyle}, {Driessen}, {Goupy}, {Kramer}, {Lagache}, {Leclercq},
  {Lestrade}, {Mac{\'\i}as-P{\'e}rez}, {Mauskopf}, {Monfardini}, {Perotto},
  {Pisano}, {Pointecouteau}, {Ponthieu}, {Rev{\'e}ret}, {Ritacco}, {Romero},
  {Roussel}, {Schuster}, {Sievers}, {Tucker}, \& {Zylka}}]{2018A&A...615A.112R}
{Ruppin}, F., {Mayet}, F., {Pratt}, G.~W., {et~al.} 2018, \aap, 615, A112

\bibitem[{{Sanders} {et~al.}(2020){Sanders}, {Dennerl}, {Russell}, {Eckert},
  {Pinto}, {Fabian}, {Walker}, {Tamura}, {ZuHone}, \&
  {Hofmann}}]{2020A&A...633A..42S}
{Sanders}, J.~S., {Dennerl}, K., {Russell}, H.~R., {et~al.} 2020, \aap, 633,
  A42

\bibitem[{{Sartoris} {et~al.}(2016){Sartoris}, {Biviano}, {Fedeli}, {Bartlett},
  {Borgani}, {Costanzi}, {Giocoli}, {Moscardini}, {Weller}, {Ascaso},
  {Bardelli}, {Maurogordato}, \& {Viana}}]{2016MNRAS.459.1764S}
{Sartoris}, B., {Biviano}, A., {Fedeli}, C., {et~al.} 2016, \mnras, 459, 1764

\bibitem[{{Sayers} {et~al.}(2023){Sayers}, {Mantz}, {Rasia}, {Allen}, {Cui},
  {Golwala}, {Morris}, \& {Wan}}]{2023ApJ...944..221S}
{Sayers}, J., {Mantz}, A.~B., {Rasia}, E., {et~al.} 2023, \apj, 944, 221

\bibitem[{Staffehl {et~al.}(2025)Staffehl, Nelson, Ayromlou, Rohr, \&
  Pillepich}]{staffehl2025abundanceorigincoolgas}
Staffehl, M., Nelson, D., Ayromlou, M., Rohr, E., \& Pillepich, A. 2025, The
  abundance and origin of cool gas in galaxy clusters in the TNG-Cluster
  simulation

\bibitem[{{Sunyaev} \& {Zeldovich}(1980)}]{1980ARA&A..18..537S}
{Sunyaev}, R.~A. \& {Zeldovich}, I.~B. 1980, \araa, 18, 537

\bibitem[{{Sunyaev} \& {Zeldovich}(1972)}]{1972CoASP...4..173S}
{Sunyaev}, R.~A. \& {Zeldovich}, Y.~B. 1972, Comments on Astrophysics and Space
  Physics, 4, 173

\bibitem[{{Tazzari} {et~al.}(2018){Tazzari}, {Beaujean}, \&
  {Testi}}]{2018MNRAS.476.4527T}
{Tazzari}, M., {Beaujean}, F., \& {Testi}, L. 2018, \mnras, 476, 4527

\bibitem[{{Tumlinson} {et~al.}(2017){Tumlinson}, {Peeples}, \&
  {Werk}}]{2017ARA&A..55..389T}
{Tumlinson}, J., {Peeples}, M.~S., \& {Werk}, J.~K. 2017, \araa, 55, 389

\bibitem[{{Ubertosi} {et~al.}(2023){Ubertosi}, {Gitti}, \&
  {Brighenti}}]{2023A&A...670A..23U}
{Ubertosi}, F., {Gitti}, M., \& {Brighenti}, F. 2023, \aap, 670, A23

\bibitem[{{van Marrewijk} {et~al.}(2024){van Marrewijk}, {Di Mascolo}, {Gill},
  {Battaglia}, {Battistelli}, {Bond}, {Devlin}, {Doze}, {Dunkley}, {Knowles},
  {Hincks}, {Hughes}, {Hilton}, {Moodley}, {Mroczkowski}, {Naess}, {Partridge},
  {Popping}, {Sif{\'o}n}, {Staggs}, \& {Wollack}}]{2023arXiv231006120V}
{van Marrewijk}, J., {Di Mascolo}, L., {Gill}, A.~S., {et~al.} 2024, \aap, 689,
  A41

\bibitem[{{Watson} {et~al.}(2003){Watson}, {Carreira}, {Cleary}, {Davies},
  {Davis}, {Dickinson}, {Grainge}, {Guti{\'e}rrez}, {Hobson}, {Jones},
  {Kneissl}, {Lasenby}, {Maisinger}, {Pooley}, {Rebolo}, {Rubi{\~n}o-Martin},
  {Rusholme}, {Saunders}, {Savage}, {Scott}, {Slosar}, {Sosa Molina}, {Taylor},
  {Titterington}, {Waldram}, \& {Wilkinson}}]{vsa}
{Watson}, R.~A., {Carreira}, P., {Cleary}, K., {et~al.} 2003, MNRAS, 341, 1057

\bibitem[{{White} {et~al.}(1997){White}, {Becker}, {Helfand}, \&
  {Gregg}}]{1997ApJ...475..479W}
{White}, R.~L., {Becker}, R.~H., {Helfand}, D.~J., \& {Gregg}, M.~D. 1997,
  \apj, 475, 479

\bibitem[{{XRISM Collaboration} {et~al.}(2025){XRISM Collaboration}, {Audard},
  {Awaki}, {Ballhausen}, {Bamba}, {Behar}, {Boissay-Malaquin}, {Brenneman},
  {Brown}, {Corrales}, {Costantini}, {Cumbee}, {Done}, {Dotani}, {Ebisawa},
  {Eckart}, {Eckert}, {Enoto}, {Eguchi}, {Ezoe}, {Foster}, {Fujimoto},
  {Fujita}, {Fukazawa}, {Fukushima}, {Furuzawa}, {Gallo}, {Garc{\'\i}a}, {Gu},
  {Guainazzi}, {Hagino}, {Hamaguchi}, {Hatsukade}, {Hayashi}, {Hayashi},
  {Hell}, {Hodges-Kluck}, {Hornschemeier}, {Ichinohe}, {Ishida}, {Ishikawa},
  {Ishisaki}, {Kaastra}, {Kallman}, {Kara}, {Katsuda}, {Kanemaru}, {Kelley},
  {Kilbourne}, {Kitamoto}, {Kobayashi}, {Kohmura}, {Kubota}, {Leutenegger},
  {Loewenstein}, {Maeda}, {Markevitch}, {Matsumoto}, {Matsushita}, {McCammon},
  {McNamara}, {Mernier}, {Miller}, {Miller}, {Mitsuishi}, {Mizumoto}, {Mizuno},
  {Mori}, {Mukai}, {Murakami}, {Mushotzky}, {Nakajima}, {Nakazawa}, {Ness},
  {Nobukawa}, {Nobukawa}, {Noda}, {Odaka}, {Ogawa}, {Ogorzalek}, {Okajima},
  {Ota}, {Paltani}, {Petre}, {Plucinsky}, {Porter}, {Pottschmidt}, {Sato},
  {Sato}, {Sawada}, {Seta}, {Shidatsu}, {Simionescu}, {Smith}, {Suzuki},
  {Szymkowiak}, {Takahashi}, {Takeo}, {Tamagawa}, {Tamura}, {Tanaka},
  {Tanimoto}, {Tashiro}, {Terada}, {Terashima}, {Trigo}, {Tsuboi}, {Tsujimoto},
  {Tsunemi}, {Tsuru}, {Uchida}, {Uchida}, {Uchida}, {Uchiyama}, {Ueda}, {Uno},
  {Vink}, {Watanabe}, {Williams}, {Yamada}, {Yamada}, {Yamaguchi}, {Yamaoka},
  {Yamasaki}, {Yamauchi}, {Yamauchi}, {Yaqoob}, {Yoneyama}, {Yoshida},
  {Yukita}, {Zhuravleva}, {Kondo}, {Werner}, {Pl{\v{s}}ek}, {Sun}, {Hosogi}, \&
  {Majumder}}]{2025Natur.638..365X}
{XRISM Collaboration}, {Audard}, M., {Awaki}, H., {et~al.} 2025, \nat, 638, 365

\bibitem[{{Zitrin} {et~al.}(2015){Zitrin}, {Fabris}, {Merten}, {Melchior},
  {Meneghetti}, {Koekemoer}, {Coe}, {Maturi}, {Bartelmann}, {Postman},
  {Umetsu}, {Seidel}, {Sendra}, {Broadhurst}, {Balestra}, {Biviano}, {Grillo},
  {Mercurio}, {Nonino}, {Rosati}, {Bradley}, {Carrasco}, {Donahue}, {Ford},
  {Frye}, \& {Moustakas}}]{2015ApJ...801...44Z}
{Zitrin}, A., {Fabris}, A., {Merten}, J., {et~al.} 2015, \apj, 801, 44

\bibitem[{{ZuHone} {et~al.}(2019){ZuHone}, {Zavala}, \&
  {Vogelsberger}}]{2019ApJ...882..119Z}
{ZuHone}, J.~A., {Zavala}, J., \& {Vogelsberger}, M. 2019, \apj, 882, 119

\bibitem[{{Zwart} {et~al.}(2008){Zwart}, {Barker}, {Biddulph}, {Bly}, {Boysen},
  {Brown}, {Clementson}, {Crofts}, {Culverhouse}, {Czeres}, {Dace}, {Davies},
  {D'Alessandro}, {Doherty}, {Duggan}, {Ely}, {Felvus}, {Feroz}, {Flynn},
  {Franzen}, {Geisb{\"u}sch}, {G{\'e}nova-Santos}, {Grainge}, {Grainger},
  {Hammett}, {Hills}, {Hobson}, {Holler}, {Hurley-Walker}, {Jilley}, {Jones},
  {Kaneko}, {Kneissl}, {Lancaster}, {Lasenby}, {Marshall}, {Newton}, {Norris},
  {Northrop}, {Odell}, {Petencin}, {Pober}, {Pooley}, {Pospieszalski}, {Quy},
  {Rodr{\'\i}guez-Gonz{\'a}lvez}, {Saunders}, {Scaife}, {Schofield}, {Scott},
  {Shaw}, {Shimwell}, {Smith}, {Taylor}, {Titterington}, {Veli{\'c}},
  {Waldram}, {West}, {Wood}, {Yassin}, \& {AMI Consortium}}]{zwart2008}
{Zwart}, J.~T.~L., {Barker}, R.~W., {Biddulph}, P., {et~al.} 2008, MNRAS, 391,
  1545

\end{thebibliography}

\begin{appendix}

\section{Spectral analysis}
\label{sec:pointsource}

\begin{figure}[h]
    \centering
    \includegraphics[scale=0.48, trim={0cm 0.cm 0cm 0cm}]{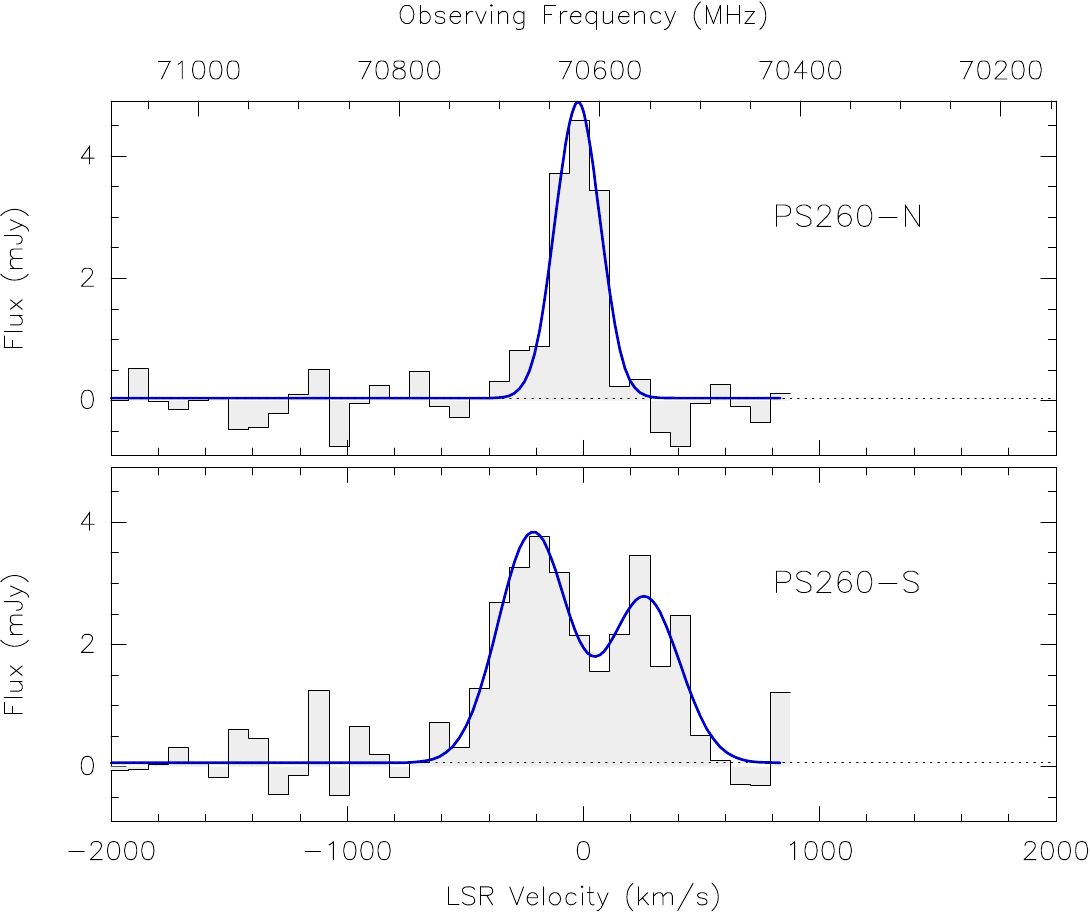}
    \caption{
    Primary beam corrected spectra towards $\text{PS260-N}$ (top) and $\text{PS260-S}$ (bottom). Velocities are calculated relative to the central frequency, determined as the average of the three Gaussian peak frequencies. The continuum has not been subtracted.
    }
    \label{fig:lines}
\end{figure}

From visual inspection of the spectral data, we identified two nearby line emission regions ($\text{PS260-S}$ and $\text{PS260-N}$) separated by 6.9$^{\prime\prime}$ (54 kpc) in projection. As shown in Fig.~\ref{fig:lines}, the two primary beam corrected spectra are centred at 70.622~GHz and 70.611~GHz, with the former ($\text{PS260-N}$) showing a narrow Gaussian profile with a FWHM of 220 km/s and an intensity of 1.1 Jy km/s, and the latter ($\text{PS260-S}$) a broad double Gaussian profile with a FWHM of 820 km/s and an intensity of 2.4 Jy km/s.
The spatial proximity of the two regions and the fact that the lines are likely at the same redshift suggest that they originate from a binary system or possibly a triple system, given the broad profile of $\text{PS260-S}$, potentially a merging system of galaxies.

The emission centroid of the two line emission regions is located at (RA, Dec)$_{\mathrm{J2000}}$ = (12$^{\rm h}$27$^{\rm m}$00$^{\rm s}$, +33$^\circ$32$'$37.5$^{\prime\prime}$), approximately $27^{\prime\prime}$ southeast of the cluster's core. The northern region ($\text{PS260-N}$), which is centred at 
(RA, Dec)$_{\mathrm{J2000}}$ = (12$^{\rm h}$27$^{\rm m}$00.02$^{\rm s}$, +33$^\circ$32$'$43.7$^{\prime\prime}$)
shows, within the uncertainties, positional coincidence with PS260 \citep{2015A&A...576A..12A}, a sub-millimetre galaxy also known as PS1 \citep{2023A&A...671A..28M} and which was previously detected by NIKA and NIKA2 \citep{2015A&A...576A..12A,2023A&A...671A..28M}. 
The two lines are likely linked to a low-J CO transition.
NOEMA might have detected the $J$=2-1 transition, which would mean that the northern source is at a redshift of $z = 2.264$. 

$\text{PS260-N}$ shows also positional coincidence (separated by 1.42$^{\prime\prime}$) with the `CL1226 1.1' gravitationally lensed source reported in \citet{2015ApJ...801...44Z}, to which a photometric redshift of $z = 2.04$ was attributed. This is also the closest galaxy to $\text{PS260-S}$ observed at other wavebands.

\section{End-to-end simulation}
\label{sec:simulationfit}

We verified the reliability of the pressure model fitting procedure presented in Sect.~\ref{sec:fitpress} via an end-to-end simulation.

\begin{figure}
    \centering
    \includegraphics[scale=0.44, trim={0.6cm 0cm 0cm 0cm}]{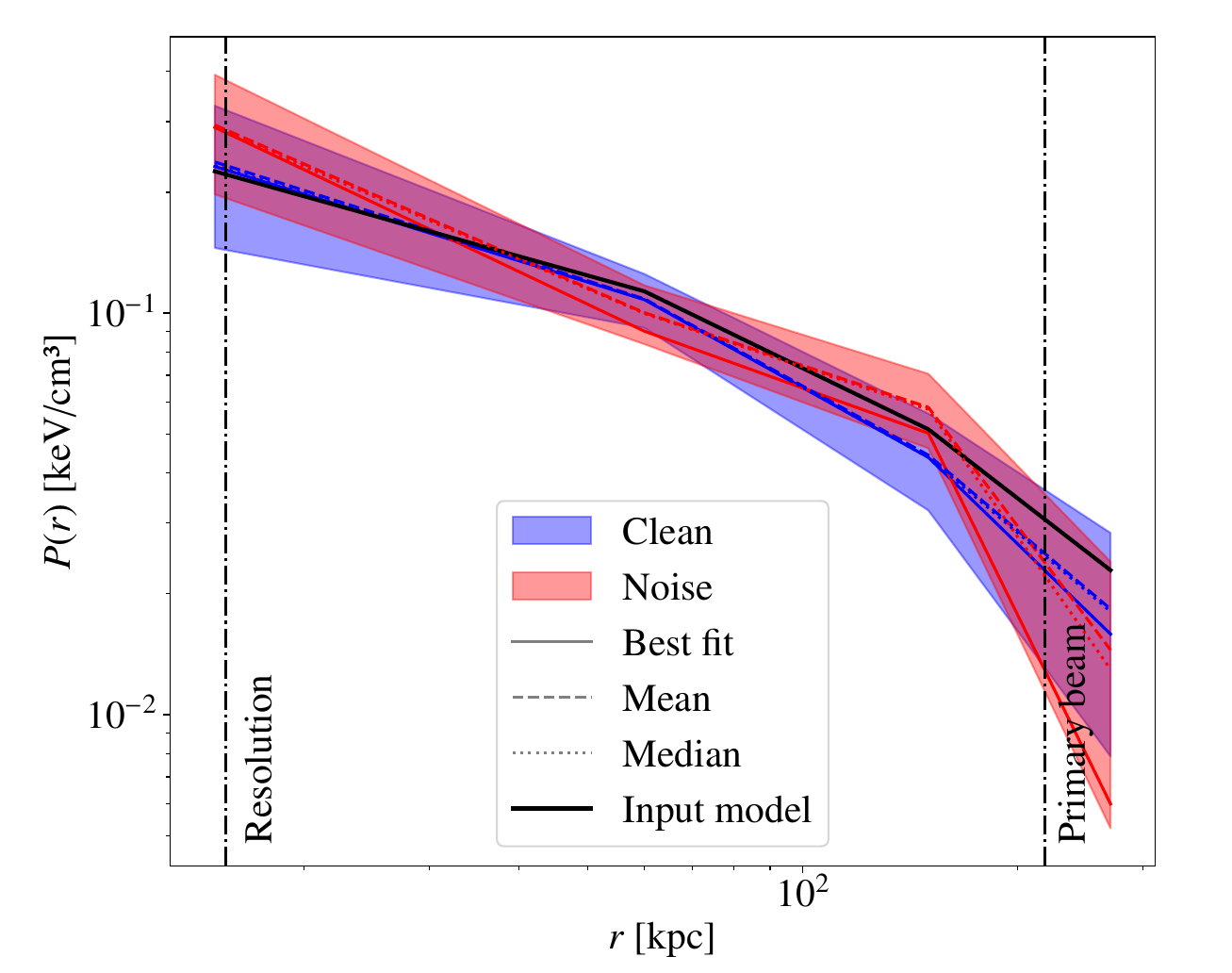}
    \caption{Pressure profiles fitted to simulated visibilities. Blue and red shaded area indicate the 16th to 84th percentiles for pressure models fitted to "Clean" and "Noise" $uv$ tables, respectively. Solid, dashed, and dotted profiles correspond respectively to the best-fit, mean, and median values of the posterior distributions of the fitted pressure bins. The mock pressure profile used as an input model for the end-to-end test is shown in black. Vertical lines indicate the angular scales accessible to the NOEMA data.
    }
    \label{fig:endtoendpressure}
\end{figure}

First, a tSZ surface brightness map was created at 82 GHz from the integration along the line of sight of the pressure profile shown in black in Fig.~\ref{fig:endtoendpressure} (a binned profile sampled at 4 radial distances), assuming a redshift of $z=0.89$. Then, the simulated image was corrected for the primary beam attenuation expected at the central frequency of each of the four basebands (i.e., four differently attenuated images were obtained this way, one for each baseband). The amplitudes of the primary beam attenuated images were then corrected for the difference in tSZ signal amplitude between the reference frequency of the simulated image (82 GHz) and the central frequency of the corresponding basebands.

For each primary beam and amplitude corrected image, a $uv$ table that fills the $uv$ plane at exactly the same locations as was done during the observations was generated. Therefore, four $uv$ tables were produced. These four $uv$ tables were then merged to one, the same way it was done for the observed data (as described in Sect.~\ref{sec:continuum}). This is the "Clean" $uv$ table. In addition, Gaussian noise was added to each visibility of each of the four $uv$ tables according to the system temperatures in the observed visibility data. The four noisy $uv$ tables were also merged to one: the "Noise" $uv$ table.

\begin{figure}
    \centering
    \includegraphics[scale=0.6]{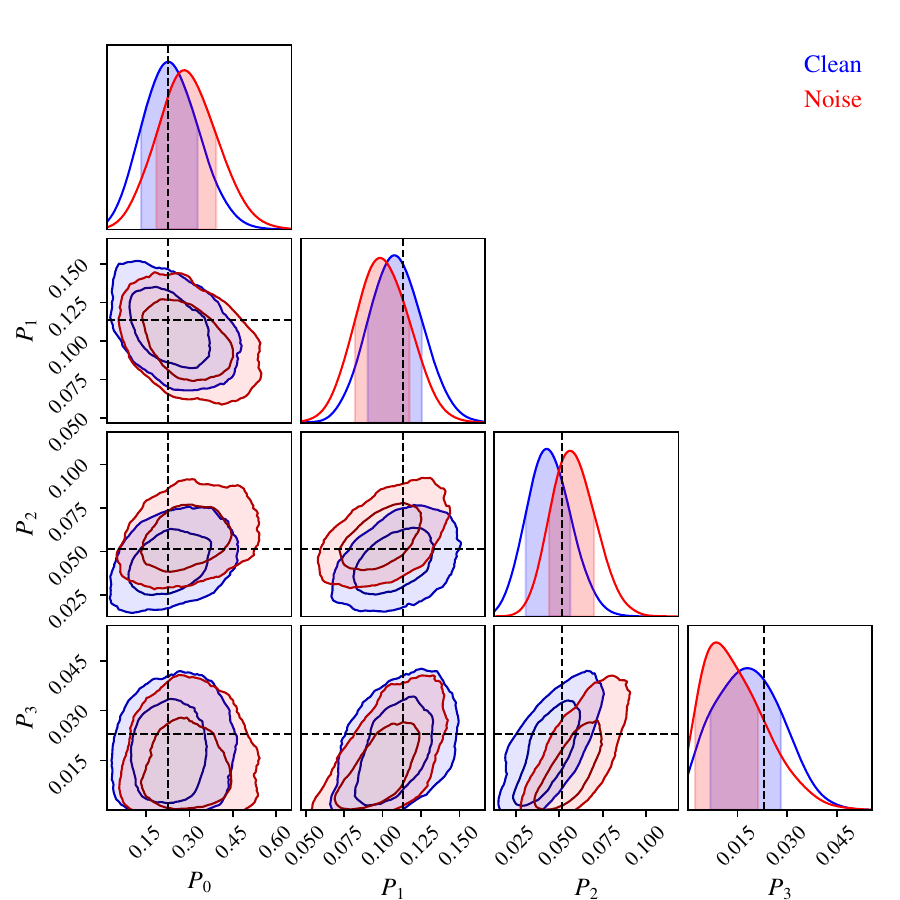}
    \caption{Posterior probability distributions (1D and 2D) of the pressure bins fitted to the simulated "Clean" (blue) and "Noise" (red) $uv$ tables. Dashed black lines indicate the values for the input model. Values are given in keV/cm$^3$ units.}
    \label{fig:endtoendcorner}
\end{figure}

The visibilities in both the "Clean" and "Noise" simulated $uv$ tables were fitted with a power law pressure model (at 15, 60, 150, and 270~kpc) following the steps described in Sect.~\ref{sec:fitpress}. Blue and red contours in Fig.~\ref{fig:endtoendcorner} show the posterior distributions of the fitted pressure bin values for the "Clean" and "Noise" $uv$ tables, respectively. Dashed black lines indicate the pressure values for each radial bin in the input model. Similarly, we present the 16th to 84th percentiles of the fitted pressure models with blue and red shaded profiles in Fig.~\ref{fig:endtoendpressure}. We verify that the recovered pressure profiles are compatible with the profile used as an input model within the angular scales accessible to the NOEMA data. 

\begin{figure}
    \centering
    \includegraphics[scale=0.37, trim={1cm 0cm 0cm 0cm}]{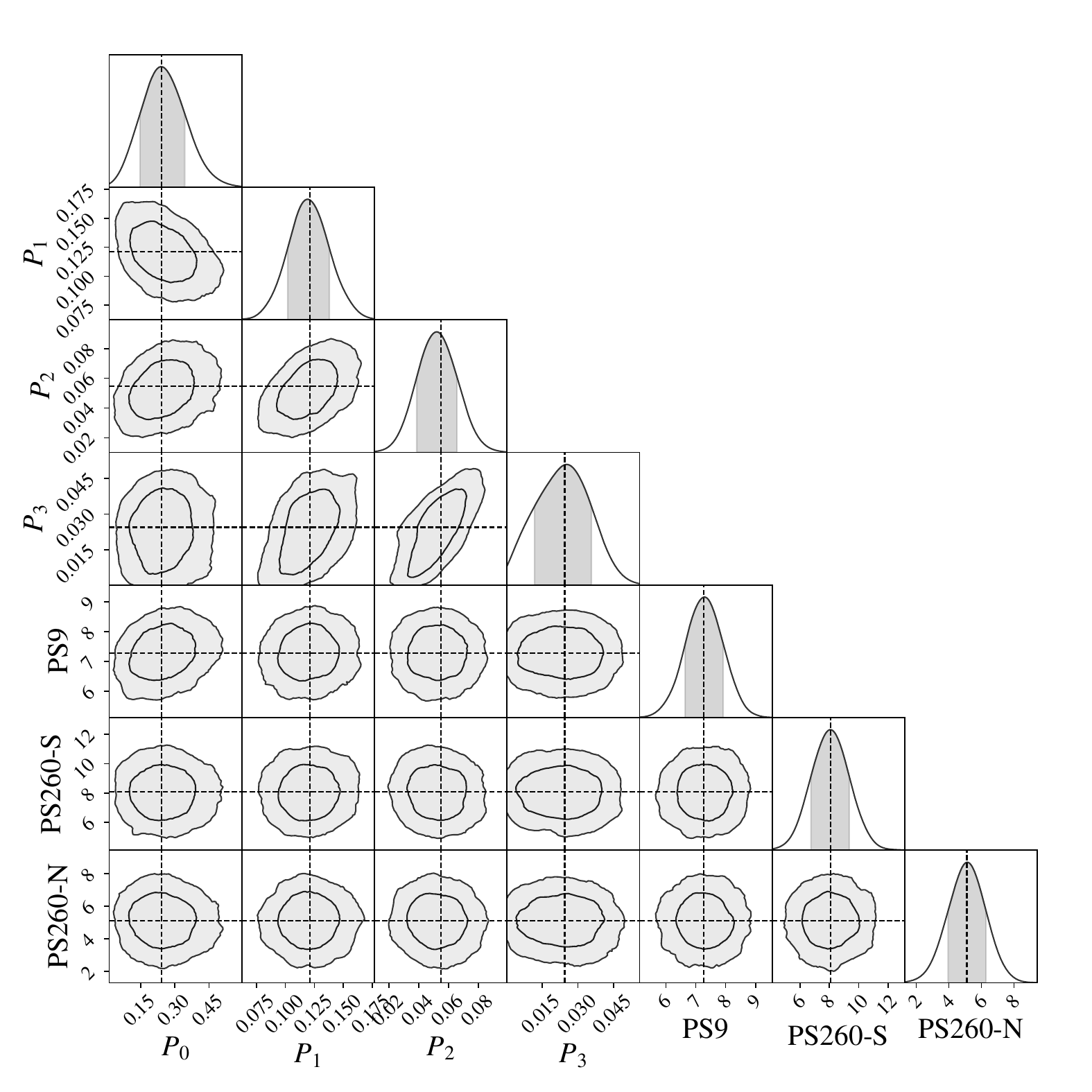}
    \caption{Posterior probability distributions (1D and 2D) of the pressure bins and point source fluxes fitted to mock visibilities. Dashed black lines indicate parameter values for the input model. Pressure values are given in keV/cm$^3$ units and point source fluxes in $10^{-5}$ Jy.}
    \label{fig:mock_psfit_test}
\end{figure}

In parallel, we tested the ability to jointly fit point source fluxes and pressure profile bins. We created again a tSZ surface brightness map at 82~GHz and added three point sources at the positions of the sources detected in our NOEMA map (Table~\ref{tab:firstfit}). Once attenuated by the primary beam, from this map we generated a $uv$ table filling the $uv$ plane at the locations of NOEMA data. Following the steps described in Sect.~\ref{sec:fitpress}, we fitted the power law pressure model (at 15, 60, 150, and 270~kpc) and point source fluxes to these mock visibilities. In Fig.~\ref{fig:mock_psfit_test}, we demonstrate that the posterior distributions of the fitted parameters (in grey) are fully consistent with the input values (dashed black lines).

\end{appendix}

\end{document}